\documentclass[fleqn,usenatbib]{mnras}

\usepackage[T1]{fontenc}
\usepackage{ae,aecompl}

\usepackage{graphicx}	
\usepackage{amsmath}	
\usepackage{amssymb}	
\usepackage{pdflscape}	
\usepackage{booktabs}
\usepackage{multirow}

\usepackage{color}

\title[Galaxy hosts of luminous AGN at  $z \sim$ 0.3-0.4]{The host galaxies of luminous  type 2 AGN at  $z \sim$0.3-0.4}
\author[Urbano Mayorgas et al.]{J.J. Urbano-Mayorgas$^{1}$\thanks{E-mail:jurbano@cab.inta-csic.es}, M. Villar Mart\'{i}n$^{1}$, F. Buitrago$^2$,   J. Piqueras L\'opez$^1$ \\
\newauthor  B. Rodr\'iguez del Pino$^1$, A. M. Koekemoer$^3$,  M. Huertas-Company$^{4,5}$ \\
\newauthor R. Dom\'inguez-Tenreiro$^{6,7}$, F.J. Carrera$^8$, C. Tadhunter$^9$ \\
$^1$Centro de Astrobiolog\'{i}a (CSIC-INTA), Carretera de Ajalvir, km 4, 28850 Torrej\'on de Ardoz, Madrid, Spain\\
$^2$Instituto de Astrof\'{i}sica e Ci\^encias do Espa\c{c}o, Universidade do Lisboa, Tapada da Ajuda - Edificio Leste 2 Piso, 1349-018, Lisboa, Portugal  \\
$^3$Space Telescope Science Institute, 3700 San Martin Drive, Baltimore, MD 21218, USA \\
$^4$Sorbonne Universit\'e, Observatoire de Paris, Universit\'e PSL, CNRS, LERMA, F-75014, Paris, France\\ 
$^5$Sorbonne Paris Cit\'e, Universit\'e Paris Diderot, 5 Rue Thomas Mann, F-75013, France \\
$^6$Dept. de F\'isica Te\'orica, Univ. Aut\'onoma de Madrid, E-28049 Cantoblanco Madrid, Spain \\
$^7$Astro-UAM, UAM, Unidad Asociada CSIC, E-28049 Cantoblanco, Madrid, Spain \\
$^8$Instituto de F\'\i sica de Cantabria (CSIC-UC), Avenida de los Castros, E-39005 Santander, Spain\\
$^9$Department of Physics and Astronomy, University of Sheffield, Sheffield S3 7RH, UK}

\begin{document}

\date{Accepted ?.
      Received ?;
      in original form ?.}

\pagerange{\pageref{firstpage}--\pageref{lastpage}}
\pubyear{2018}

\maketitle

\label{firstpage}

\begin{abstract}

We study the morphological and structural properties of the host galaxies associated with 57 optically-selected luminous type 2 AGN at $z\sim$0.3-0.4: 16 high-luminosity Seyfert 2 (HLSy2, 8.0$\le$log($L_{\rm [OIII]}/L_{\odot})<$8.3) and 41 obscured quasars (QSO2, log($L_{\rm [OIII]}/L_{\odot})\ge$8.3). With this work, the total number of QSO2 at $z<1$ with parametrized galaxies increases from $\sim$35 to 76. Our analysis is based on  HST WFPC2 and ACS images that we fit with {\sc GALFIT}.  HLSy2 and QSO2 show a wide diversity of galaxy hosts.  The main difference lies in the higher incidence of highly-disturbed systems among QSO2. This is consistent with a scenario in which galaxy interactions are the dominant mechanism triggering nuclear activity at the highest AGN power. There is a strong dependence of galaxy properties with AGN power (assuming $L_ {\rm [OIII]}$ is an adequate proxy). The relative contribution of the spheroidal  component  to the total galaxy light (B/T) increases with $L_ {\rm [OIII]}$. While systems dominated by the spheoridal component spread across the total range of $L_ {\rm [OIII]}$, most disk-dominated galaxies concentrate at log($L_{\rm [OIII]}/L_{\odot})<$8.6. This is expected if more powerful AGN are powered by more massive black holes which are hosted by more massive bulges or spheroids.  The average galaxy sizes ($\langle r_{\rm e} \rangle$) are 5.0$\pm$1.5 kpc for HLSy2  and 3.9$\pm$0.6 kpc for HLSy2 and QSO2 respectively. These are significantly smaller than those found for QSO1 and narrow line radio galaxies at similar $z$. We put the results of our work in context of related studies of AGN with quasar-like luminosities.

\end{abstract}

\begin{keywords}

galaxies: active - galaxies: evolution - galaxies:quasars:general 

\end{keywords}

\section{Introduction}
\label{Sec:intro}

Studies of the host galaxies associated with active galactic nuclei (AGN)  are relevant  on a diversity of topics related to galaxy formation and evolution such as:  
What mechanisms control nuclear activity and supermassive black hole (SMBH) growth in galaxies? What is the role of orientation and obscuration in the observed differences among certain AGN sub-classes? How is radio activity triggered?   What is the origin of the tight scaling relations between the  SMBH masses and various properties of their host spheroids? Ultimately,  what is the link between galaxy and SMBH formation and evolution?

Quasars are the most powerful active galaxies. By studying their host galaxies at different redshift ($z$)  we can investigate how the most massive black holes form and evolve, what mechanisms trigger the most extreme form of nuclear activity and how this can affect the evolution  of massive galaxies  \citep{Kormendi1995, Magorrian1998,Ferrarese2000, Gebhardt2000, Tremaine2002}. Such studies have been  focussed traditionally on type 1 (unobscured) quasars (QSO1). While some works proposed that   QSO1 at  low redshift (z$<$0.5)  are almost invariably hosted by  massive, bulge-dominated galaxies \citep{McLeod1994,McLeod1995b,Dunlop2003,Lacy2006,Hyvonen2007}, other studies have shown a large diversity of  hosts.  A  substantial disk component has been found in many galaxies hosting low-$z$ quasars, with relative contribution to the total galaxy light possibly dependent on quasar luminosity and radio-loudness (\citealp{Bahcall1997,Jahnke2004,Floyd2004,Bettoni2015}).

In regard to the physical mechanisms that trigger  AGN activity and SMBH growth, there is evidence supporting that a variety of processes can be involved, with the dominant one  depending on  AGN luminosity. While mergers of gas-rich galaxies are frequently suggested as the trigger for quasars, secular processes  appear to be more relevant at lower AGN power \citep{Toomre1972,Heckman1986,Combes2001,Hopkins2006,Cisternas2011,CRamos2011,Bessiere2012}. 

Host galaxy studies of type 2  (obscured)	QSO (QSO2) at different $z$ are currently of special relevance.  This population is at least comparable in number density to the QSO1 population and perhaps 2-3 times larger \citep{Tajer2007,Gili2011,Mateos2017}. They are of great interest since they are signposts of vigorous obscured SMBH growth. 

In comparison with QSO1 studies they can provide useful information regarding the QSO1 versus QSO2 unification scenario based on orientation \citep{Antonucci1993}. 

Only $\sim$10-15\%  of quasars are radio-loud. This applies to both QSO1 \citep{Katgert1973,Fanti1977,Smith1980} and QSO2 \citep{Lal2010}.  QSO2 studies offer the opportunity   to characterize the host galaxies of the most luminous obscured radio-quiet AGN versus  their radio-loud analogues, narrow line radio galaxies (NLRG, e.g. \citealt{Dunlop2003,Best2005,Inskip2010}).

QSO2 have  been discovered in large numbers  only recently \citep{Zakamska2003}.  For this reason, studies of their hosts  are scarce and have been focussed on small samples. Such studies  have a clear advantage with respect to QSO1: the obscuration of the central engine  renders  a  detailed view of the galaxies, allowing  a more accurate morphological and structural characterization. 
These works suggest a diversity of galaxy host types,  with a clear preference for ellipticals and bulge-dominated systems   (\citealt{Greene2009},  \citealt{Bessiere2012}, \citealt{VillarM2012}, \citealt{Kocevski2012}, \citealt{Wylezalek2016}). 

With the goal of shedding more light on this topic, we present here the results of the morphological and parametric characterization and subsequent classification of the host galaxies associated with 57 luminous obscured AGN  at z $\sim$ 0.3-0.4. 41 are QSO2.  In order to investigate the potential dependence of galaxy host properties with AGN power,  16 high-luminosity Seyfert 2 galaxies (HLSy2) are also part of this study (\citealt{McLeod1995a}, \citealt{Kauffmann2003}).

We also identify and classify  merger/interaction features.     
 Our study is based on  Hubble Space Telescope (HST) optical images  obtained with the  Advanced Camera for Surveys/Wide Field Channel (ACS/WFC) and the Wide Field Planetary Camera 2 (WFPC2).
We have applied  two different techniques: a visual classification and  multi-parametric modelling, using {\sc GALFIT} \citep{Peng2010}, which allows to isolate and parametrize the galaxy structural components.

The paper is organised as follows. The AGN sample and data are described in Section \ref{Sample:data}. The classification methods and the modelling procedure are  explained in Section \ref{Class:methods}. The results of the visual and parametric classifications are  presented in Section \ref{Results} and  discussed in the context of related works in Section \ref{Discusion}. Summary and conclusions are in Section \ref{Conclusions}.

We assume   $\Omega_\Lambda=0.73$ , 	$\Omega_M=0.27$ and $H_0=$71 km s$^{-1}$ Mpc$^{-1}$.

\section{Sample and data}
\label{Sample:data}

The sample studied here consists  of 57 luminous ($lO3$=log($L_{\rm [OIII]}/L_{\odot})\ge$8.0) type 2 AGN at $0.3 < z < 0.4$  from   \citet{Zakamska2003} \& \citet{Reyes2008} catalogues of  Sloan Digital Sky Survey (SDSS) luminous type 2 AGN  (Table \ref{Table:0}).

 \citet{Zakamska2003} selected  291 luminous type 2 AGN  ($lO3>$7.3), at $z<0.83$ from SDSS  on the basis of their optical emission line properties: narrow emission lines (FWHM$<$2000 km s$^{-1}$) without underlying broad components, and optical line ratios typical of active galaxies, consistent with non-stellar ionizing radiation.   \citet{Reyes2008} updated  this  catalogue based on  $\sim$3 times as much SDSS data. Their catalogue contains 887 luminous type 2 AGN ($lO3>$7.9), recovering $>$90\% of objects in \citet{Zakamska2003} in the same luminosity range. The spectra of the objects they missed tend to have low S/N or ambiguous classification.  

About 744 (84\%)  objects in  \citet{Reyes2008}  have  $lO3\ge$8.3 and are, therefore, QSO2. This threshold ensures the selection of objects with AGN luminosities in the quasar regime. Using $L_ {\rm [OIII]}$  as a proxy for AGN power \citep{Heckman2004},  the implied  bolometric luminosities are above the classical Seyfert/quasar separation of $L_{\rm bol}\sim$10$^{45}$ erg s$^{-1}$. Only $\sim$15\%$\pm$5\%  QSO2 are expected to be radio-loud \citep{Lal2010}.

The 57 AGN studied here are the sample of objects observed for the    Hubble Space Telescope (HST)  program  10880, with principal investigator Henrique Schmitt (Tables \ref{Table:0} and  \ref{Table:0b}). HST imaging observations for other programmes exist for several more QSO2, but  in general  they have been done with different filters and/or the targets are at different $z$ than our sample. Since the statistics will not improve significantly, these are not considered in our study.

There are  97 SDSS  QSO2 and 36 HLSy2 in the $0.3 < z < 0.4$ range. Of these, our subsample contains 
41 ($\sim$42\%)  QSO2 and 16 HLSy2  ($\sim$44\%).  Although uncertainties remain regarding the exact selection criteria applied by the team responsible for the 10880 HST
program, based on the high fractions quoted above we consider they are an adequate representation of the total sample of SDSS QSO2 and HLSy2 in these $z$ and $L_{\rm [OIII]}$ ranges.

The ACS/WFC and WFPC2  images used in this work are from the  Hubble Legacy Archive (HLA)\footnote{https://hla.stsci.edu/}.

    \section{Classification methods}
	\label{Class:methods}

We have classified the host galaxies based on two main methods: visual and parametric.

The visual inspection of host galaxy images provides a   classification based on their apparent morphology. It has been a standard practice for more than 80 years  \citep{Hubble1936} and is contributing still today  to achieve a deeper understanding of galaxy evolution (\citealt{Lintott2008}, \citealt{Nair2010}, \citealt{Willett2013}).   A limitation of this method is that it can be subjective, so that the same object can be classified differently by different observers. Frequently, it does not allow to determine which structural component (disk or bulge) dominates   the total galaxy light, thus preventing an accurate classification.  A more robust classification  needs to be based on a multi-parametric modelling approach. This  allows to extract structural components from galaxy images, by modeling their light profiles \citep{Peng2010}. This method has also limitations. As an example, complex mergers can be missclassified by applying a too simplistic approach of assuming that all galaxies consist of a disk and/or a spheroidal component. Visual inspection is particularly useful in these cases.

Both methods have been essential in our work: the visual classification has allowed to identify complex mergers that cannot be classified neither as bulge or disk dominated systems. It has also been useful  to disentangle the   parametric classification in a minority of bulge or disk dominated cases where the parametric method resulted on degenerate fits.  
	
More details on the classification methods are provided next.

	\subsection{Visual classification}
	\label{Visual:class}

We have applied three different methods of visual classification:

\begin{itemize}
\setlength\itemsep{1em}

\item Method Vis-I.    Three groups have been considered: spiral and disks with not obvious spiral arms, ellipticals and highly-disturbed systems. These are systems of very complex morphologies due to merger/interaction processes which cannot be classified in the previous two groups.  

\par
 
\item Method Vis-II focuses on the identification  of features indicative of galaxy interactions. 

Given that QSO2 host galaxies are often associated with morphological features indicative of past or ongoing merger/interaction
events \citep{VillarM2012, Bessiere2012}, we have classified our objects to highlight the presence of such features, adopting the following schemes  of \citet{Rodriguez2011} and \citet{Veilleux2002}:

\begin{itemize}
 \item Class 0: Objects that appear to be single isolated galaxies, dominated by a relatively symmetric  morphology with no peculiar features.
 
  \item Class 0*: Objects that appear to be single isolated galaxies, dominated by a  symmetric  morphology with some faint irregular morphological features such as tails, shells, etc (see Method Vis-III below).
 
 \item Class 1: objects in a pre-coalescence phase with two well differentiated nuclei separated a projected distance $>$1.5 kpc. For these objects, it is still possible to identify the individual merging galaxies and their corresponding tidal structures due to the interaction.
 
 \item Class 2: objects with two nuclei separated a projected distance $\leq$1.5 kpc or a single nucleus with an asymmetric morphology and prominent irregular features suggesting a post-coalescence merging phase.
\end{itemize}

\item Method Vis-III. Morphological appearance of peculiar features. To further refine this classification, we  have also characterized  the morphological appearance of the merger/interaction features following \citet{CRamos2011}: T: tidal tail; F: Fan; B: Bridge; S: Shell; D: dust feature; 2N: Dual-Core/Double Nucleus; A: amorphous halo; I: irregular feature; IC:  interacting companion. We have added: K: Knot, as an extra feature for the characterization.

\end{itemize}

 In this paper we present the results of all three visual methods, although we  will focus the scientific discussion on Method Vis-I.

\begin{center}
\begin{table*}
\footnotesize
\begin{tabular}{|l|c|c|c|c|c|c|c|c|c|c|c|}
\hline \hline 
\multicolumn{1}{|c|}{Object } 	& z 		& Scale 	     & $lO3$ & Instrument 		& Date 		& AGN		&		Emission lines 				\\
\multicolumn{1}{|c|}{SDSS Name } & 	& kpc/arcsec & 		& /filter 			& dd/mm/yr 	& Classification & 								\\
\multicolumn{1}{|c|}{[1]} 	& [2] 			& [3] 		& [4] 		& [5] 				& [6] 			& [7]      	     & [8]								\\ \hline
J002531.46-104022.2	&	0.303	&	4.45	&8.73	&	ACS/F775W  	&	11/09/06	&	QSO2   &	H$\alpha$, [NII] doublet	   			\\
J005515.82-004648.6	&	0.345	&	4.86	&8.15	&	WFPC2/F814W	&	18/06/07	&	HLSy2  &	H$\alpha$, [NII] doublet	   			\\
J011429.61+000036.7	&	0.389	&	5.25	&8.66	&	ACS/F775W  	&	07/08/06	&	QSO2   &	[OIII] 4959,5007	   		   		\\
J011522.19+001518.5	&	0.390	&	5.26	&8.14	&	WFPC2/F814W	&	11/06/07	&	HLSy2  &	H$\alpha$, [NII] doublet	   			\\
J014237.49+144117.9	&	0.389	&	5.25	&8.76	&	ACS/F775W  	&	11/08/06	&	QSO2   &	[OIII] 4959,5007	   		   		\\
J015911.66+143922.5	&	0.319	&	4.61	&8.56	&	ACS/F775W  	&	12/08/06	&	QSO2   &	No	  		 				   	\\
J020234.56-093921.9	&	0.302	&	4.44	&8.39	&	WFPC2/F814W	&	18/06/07	&	QSO2   &	H$\alpha$, [NII] doublet	   			\\
J021059.66-011145.5	&	0.384	&	5.21	&8.10	&	WFPC2/F814W	&	25/06/07	&	HLSy2  &	H$\alpha$, [NII] doublet	   			\\
J021758.18-001302.7	&	0.344	&	4.85	&8.55	&	ACS/F775W  	&	19/10/06	&	QSO2   &	No	   							\\
J021834.42-004610.3	&	0.372	&	5.10	&8.85	&	ACS/F775W  	&	15/09/06	&	QSO2   &	[OIII] 5007	   						\\
J022701.23+010712.3	&	0.363	&	5.02	&8.90	&	ACS/F775W  	&	06/11/06	&	QSO2   &	[OIII] 5007	   						\\
J023411.77-074538.4	&	0.310	&	4.52	&8.77	&	ACS/F775W  	&	30/10/06	&	QSO2   &	H$\alpha$, [NII] doublet	   			\\
J031946.03-001629.1	&	0.393	&	5.22	&8.24	&	WFPC2/F814W	&	25/11/08	&	HLSy2  &	H$\alpha$, [NII] doublet	   			\\
J031927.22+000014.5	&	0.385	&	5.28	&8.06	&	WFPC2/F814W	&	25/11/08	&	HLSy2  &	H$\alpha$, [NII] doublet	   			\\
J032029.78+003153.5	&	0.384	&	5.21	&8.52	&	ACS/F775W  	&	11/11/06	&	QSO2   &	[OIII] 4959, 5007	   		   		\\
J032533.33-003216.5	&	0.352	&	4.93	&9.06	&	WFPC2/F814W	&	26/11/08	&	QSO2   &	H$\alpha$, [NII] doublet	   			\\
J033310.10+000849.1	&	0.327	&	4.69	&8.13	&	WFPC2/F814W	&	25/11/08	&	HLSy2  &	H$\alpha$, [NII] doublet	   			\\
J034215.08+001010.6	&	0.348	&	4.89	&9.08	&	WFPC2/F814W	&	22/11/08	&	QSO2   &	H$\alpha$, [NII] doublet	   			\\
J040152.38-053228.7	&	0.320	&	4.62	&8.96	&	WFPC2/F814W	&	28/11/08	&	QSO2   &	H$\alpha$, [NII] doublet	   			\\
J074811.44+395238.0	&	0.372	&	5.10	&8.19	&	WFPC2/F814W	&	20/11/08	&	HLSy2  &	H$\alpha$, [NII] doublet	   			\\
J081125.81+073235.3	&	0.350	&	4.91	&8.88	&	WFPC2/F814W	&	22/11/08	&	QSO2   &	H$\alpha$, [NII] doublet	   			\\
J081330.42+320506.0	&	0.398	&	5.33	&8.83	&	ACS/F775W  	&	16/12/06	&	QSO2   &	[OIII] 4959, 5007	   				\\
J082449.27+370355.7	&	0.305	&	4.47	&8.28	&	WFPC2/F814W	&	23/11/08	&	QSO2   &	H$\alpha$, [NII] doublet	   			\\
J082527.50+202543.4	&	0.336	&	4.78	&8.88	&	WFPC2/F814W	&	20/11/08	&	QSO2   &	H$\alpha$, [NII] doublet	   			\\
J083028.14+202015.7	&	0.344	&	4.85	&8.91	&	WFPC2/F814W	&	20/11/08	&	QSO2   &	H$\alpha$, [NII] doublet	   			\\
J084041.08+383819.8	&	0.313	&	4.55	&8.47	&	WFPC2/F814W	&	21/11/08	&	QSO2   &	H$\alpha$, [NII] doublet	   			\\
J084309.86+294404.7	&	0.397	&	5.32	&9.34	&	WFPC2/F814W	&	21/11/08	&	QSO2   &	[OIII] 5007, H$\alpha$, [NII] doublet		\\
J084856.58+013647.8	&	0.350	&	4.91	&8.46	&	ACS/F775W  	&	08/10/06	&	QSO2   &	No	   							\\
J084943.82+015058.2	&	0.376	&	5.14	&8.06	&	WFPC2/F814W	&	26/11/08	&	HLSy2  &	H$\alpha$, [NII] doublet	   			\\
J090307.84+021152.2	&	0.329	&	4.71	&8.42	&	WFPC2/F814W	&	26/11/08	&	QSO2   &	H$\alpha$, [NII] doublet	   			\\
J090414.10-002144.9	&	0.353	&	4.94	&8.93	&	ACS/F775W  	&	11/12/06	&	QSO2   &	No	   							\\
J090801.32+434722.6	&	0.363	&	5.02	&8.31	&	WFPC2/F814W	&	21/11/08	&	QSO2   &	H$\alpha$, [NII] doublet	   			\\
J092318.06+010144.8	&	0.386	&	5.23	&8.94	&	WFPC2/F814W	&	21/11/08	&	QSO2   &	H$\alpha$, [NII] doublet	   			\\
J092356.44+012002.1	&	0.380	&	5.17	&8.59	&	WFPC2/F814W	&	17/11/08	&	QSO2   &	H$\alpha$, [NII] doublet	   			\\
J094209.00+570019.7	&	0.350	&	4.91	&8.31	&	WFPC2/F814W	&	22/11/08	&	QSO2   &	H$\alpha$, [NII] doublet	   			\\
J094350.92+610255.9	&	0.341	&	4.82	&8.46	&	WFPC2/F814W	&	15/04/07	&	QSO2   &	H$\alpha$, [NII] doublet	   			\\
J095629.06+573508.9	&	0.361	&	5.01	&8.38	&	WFPC2/F814W	&	26/11/08	&	QSO2   &	H$\alpha$, [NII] doublet	   			\\
J100329.86+511630.7	&	0.324	&	4.66	&8.11	&	WFPC2/F814W	&	26/11/08	&	HLSy2  &	H$\alpha$, [NII] doublet	   			\\
J103639.39+640924.7	&	0.398	&	5.33	&8.42	&	WFPC2/F814W	&	14/04/07	&	QSO2   &	[OIII] 5007, H$\alpha$, [NII] doublet		\\
J112907.09+575605.4	&	0.313	&	4.55	&9.38	&	WFPC2/F814W	&	26/11/08	&	QSO2   &	H$\alpha$, [NII] doublet	   			\\
J113710.78+573158.7	&	0.395	&	5.30	&9.61	&	WFPC2/F814W	&	26/11/08	&	QSO2   &	[OIII] 5007, H$\alpha$, [NII] doublet		\\
J133735.01-012815.7	&	0.329	&	4.71	&8.72	&	WFPC2/F814W	&	04/04/07	&	QSO2   &	H$\alpha$, [NII] doublet	   			\\
J140740.06+021748.3	&	0.309	&	4.51	&8.90	&	WFPC2/F814W	&	05/04/07	&	QSO2   &	H$\alpha$, [NII] doublet	   			\\
J143027.66-005614.9	&	0.318	&	4.60	&8.44	&	WFPC2/F814W	&	05/04/07	&	QSO2   &	H$\alpha$, [NII] doublet	   			\\
J144711.29+021136.2	&	0.386	&	5.23	&8.45	&	WFPC2/F814W	&	02/05/07	&	QSO2   &	H$\alpha$, [NII] doublet	   			\\
J150117.96+545518.3	&	0.338	&	4.79	&9.06	&	WFPC2/F814W	&	08/04/07	&	QSO2   &	H$\alpha$, [NII] doublet	   			\\
J154133.19+521200.1	&	0.311	&	4.53	&8.25	&	WFPC2/F814W	&	02/04/07	&	HLSy2  &	H$\alpha$, [NII] doublet	   			\\
J154337.81-004420.0	&	0.311	&	4.53	&8.40	&	WFPC2/F814W	&	05/05/07	&	QSO2   &	H$\alpha$, [NII] doublet	   			\\
J154613.27-000513.5	&	0.383	&	5.20	&8.18	&	WFPC2/F814W	&	03/05/07	&	HLSy2  &	H$\alpha$, [NII] doublet	   			\\
J172419.89+551058.8	&	0.365	&	5.04	&8.00	&	WFPC2/F814W	&	17/05/07	&	HLSy2  &	H$\alpha$, [NII] doublet	   			\\
J172603.09+602115.7	&	0.333	&	4.75	&8.57	&	WFPC2/F814W	&	08/04/07	&	QSO2   &	H$\alpha$, [NII] doublet	   			\\
J173938.64+544208.6	&	0.384	&	5.21	&8.42	&	WFPC2/F814W	&	10/04/07	&	QSO2   &	H$\alpha$, [NII] doublet	   			\\
J214415.61+125503.0	&	0.390	&	5.26	&8.14	&	WFPC2/F814W	&	14/05/07	&	HLSy2  &	H$\alpha$, [NII] doublet	   			\\
J215731.40+003757.1	&	0.390	&	5.26	&8.39	&	WFPC2/F814W	&	12/05/07	&	QSO2   &	H$\alpha$, [NII] doublet	   			\\
J223959.04+005138.3	&	0.384	&	5.21	&8.15	&	WFPC2/F814W	&	17/05/07	&	HLSy2  &	H$\alpha$, [NII] doublet	   			\\
J231755.35+145349.4	&	0.311	&	4.53	&8.10	&	WFPC2/F814W	&	21/05/07	&	HLSy2  &	H$\alpha$, [NII] doublet	   			\\
J231845.12-002951.4	&	0.397	&	5.32	&8.00	&	WFPC2/F814W	&	12/06/07	&	HLSy2  &	[OIII] 5007, H$\alpha$, [NII] doublet		\\ \hline
\end{tabular}
\caption{List of objects observed for the HST program 10880.   Col(3) quotes the kpc/arcsec conversion. The [OIII] luminosity in Col(4), $lO3$, is given in log and relative to the solar luminosity (Reyes et al. 2008). Objects with $lO3\gtrsim$ 8.3 are type 2 quasars (QSO2) in Col(7). Objects with lower values are classified as high-luminosity Seyfert 2s (HLSy2).  Col(5): HST instrument and filter. Col(6): Date of observation. Col(8): Emission lines contaminating the filter.}
\label{Table:0}
\end{table*}
\end{center}

\subsection{Parametric classification}
\label{Parametric:class}

The two dimensional (2D) fitting algorithm {\sc GALFIT} (version 3.0) has been used to model the galaxies.  This algorithm allows to extract structural components of  galaxies by modelling their light profiles with parametric functions  (\citealt{Peng2002}, \citeyear{Peng2010}).   The final fit for each galaxy consists of one or more components. One may be a point source (a point spread function,  PSF) and the rest are described by a S\'ersic (\citeyear{Sersic1963}) function:
 
\begin{equation}
\Sigma (r) ~=~ \Sigma_e ~e^{-\kappa[(r/r_e)^{1/n}-1]}
\end{equation}
 
   where $r_e$ is the effective radius of the galaxy, $\Sigma_e$ is the surface brightness at radius $r=r_e$, $n$ is the  S\'ersic index  and $\kappa$ is a parameter coupled to $n$ so that half of the total flux is  within $r_e$. The particular cases of $n=$4 (Vaucouleurs' law) and $n=$1 are often assumed to fit bulge and exponential disk components respectively.

{\sc GALFIT} fits the following parameters for each S\'ersic component: central position ($x,y$), integrated magnitude (\textit{MAG})\footnote{Correction for Galactic extinction has been taken into account},  $r_{\rm e}$,  $n$, axis ratio ($b/a$)  and position angle of the major axis ($PA$). The users need to start the algorithm with initial guesses for these parameters, that have to be as accurate as possible, and a value for the sky background.
Following different works (\citealt{Haussler2007}, \citealt{Buitrago2008, Buitrago2017}),  we obtain the  input parameters with  {\sc SEXTRACTOR} \citep{Sextractor1996}  except the sky background (see Sect. \ref{noise}).  Zero points are fixed in each object and they are provided in Table \ref{Table:0b}. Close neighbours were fitted using Single S\'ersic profiles simultaneously with the target galaxies to avoid contamination of the AGN host light profiles.

 \begin{center}
\begin{table*}
\footnotesize
\begin{tabular}{|c|c|c|c|c|c|c|c}
\hline 
Instrument  &      Pixel scale    &  FWHM(PSF) &  FWHM(PSF)   & Filter & $\Delta\lambda$  & $Z_{\rm p}$ & Nr. of objects\\ 
  &     arcsec pix$^{-1}$   & arcsec  &  kpc  &  & \AA\ \\
~[1]  &  [2] & [3]  &  [4] & [5]  &  [6]  & [7] & [8] \\  \hline
ACS/WFC	&	  	0.05	  & 0.12  & 0.54-0.65  & F775W & 6804-8632  & 25.65	& 12\\
WFCP2  &    0.1  & 0.25  & 1.1-1.4 & F814W & 6984-10043 & 24.21  & 45 \\  \hline
 \end{tabular}
\caption{Instrument specifications. [4]: range of PSF physical sizes spanned by the $z$ of the sample. [6]: spectral range covered by the filter.
 [7]: Zero point values for flux calibration \citep{Lucas2016}. [8]: Number of objects observed with each instrument.}
\label{Table:0b}
\end{table*}
\end{center}

{\sc GALFIT} provides $r_e$ and $MAG$ of the individual components, but not the global values for the galaxy. 
To obtain these,   elliptical isophotes  were fitted to the galaxy's 2D model  using the {\sc ellipse} task in {\sc IRAF}. To avoid  overestimations of  the total galaxy flux $F_ {\rm T}$ and of $r_{\rm e}$, the outer model isophote  was  carefully fixed to coincide with the HST image isophote for which   the flux per pixel is $>$3$\sigma$ (Section \ref{noise}).   The relative contribution of each structural component  is measured as the flux within this isophote relative to $F_{\rm T}$, $\frac{F_{\rm i}}{F_{\rm T}}$. The galaxy $r_e$ is taken as  the major axis of the model isophote that contains  $\frac{F_{\rm T}}{2}$ (Table \ref{Table:1}).

\begin{figure*}
 \centering
 Spheroidal, Bulge-dominated: SDSS J031927.22+000014.5\\
  \vspace{0.1in}
\includegraphics[width=\textwidth]{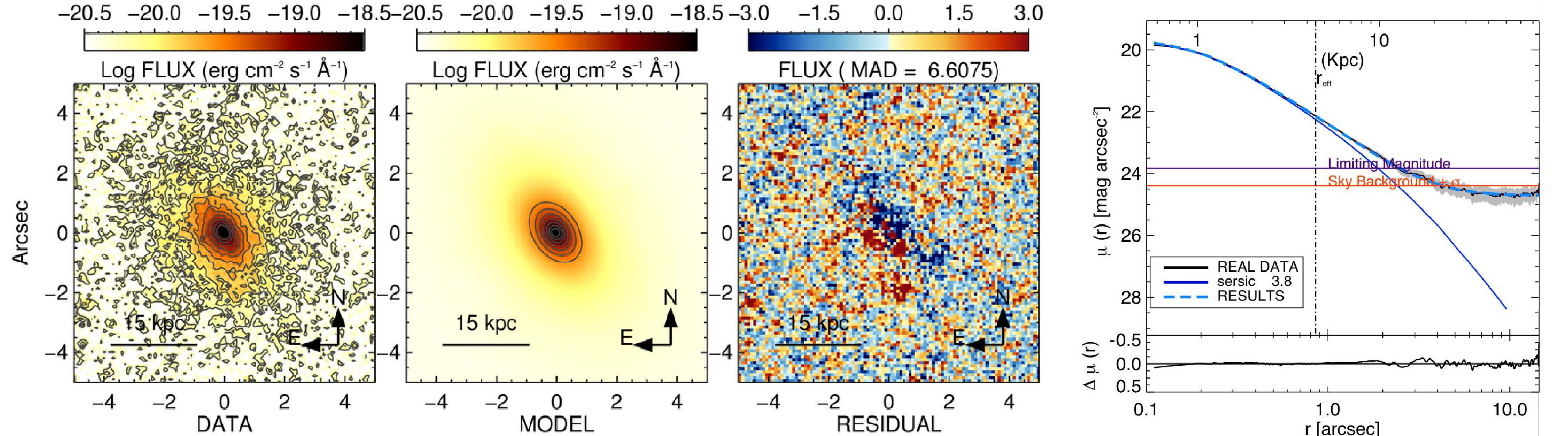} \\
 \vspace{0.1in}
 Disk-like, Disk-dominated: SDSS J005515.82-004648.6\\
  \vspace{0.1in}
 \includegraphics[width=\textwidth]{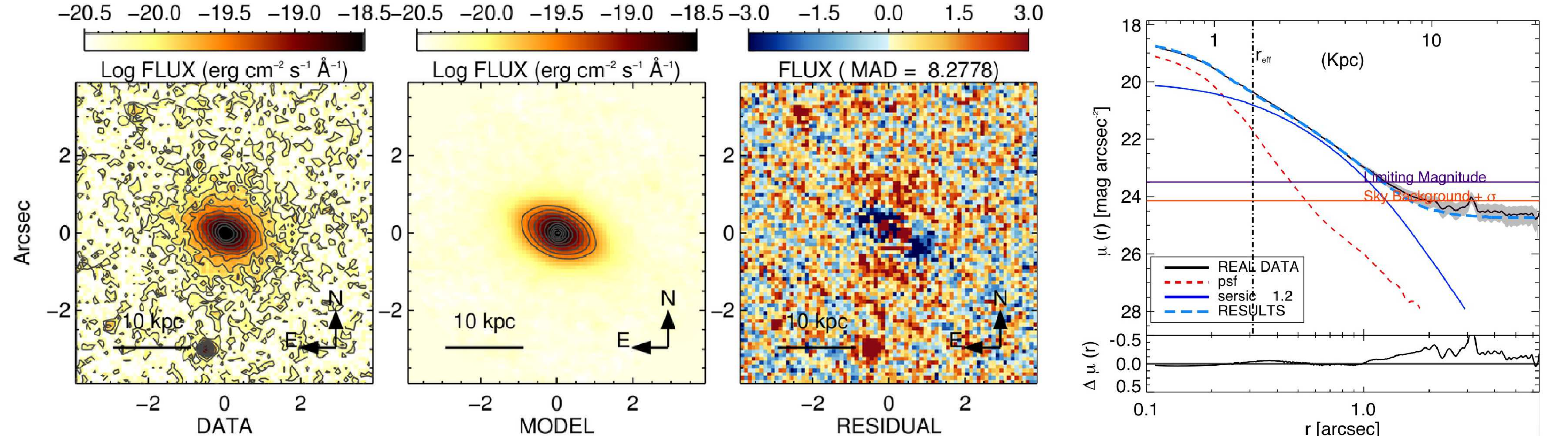} \\
  \vspace{0.1in}
Bulge-dominated: SDSS J083028.14+202015.7 \\
 \vspace{0.1in}
\includegraphics[width=\textwidth]{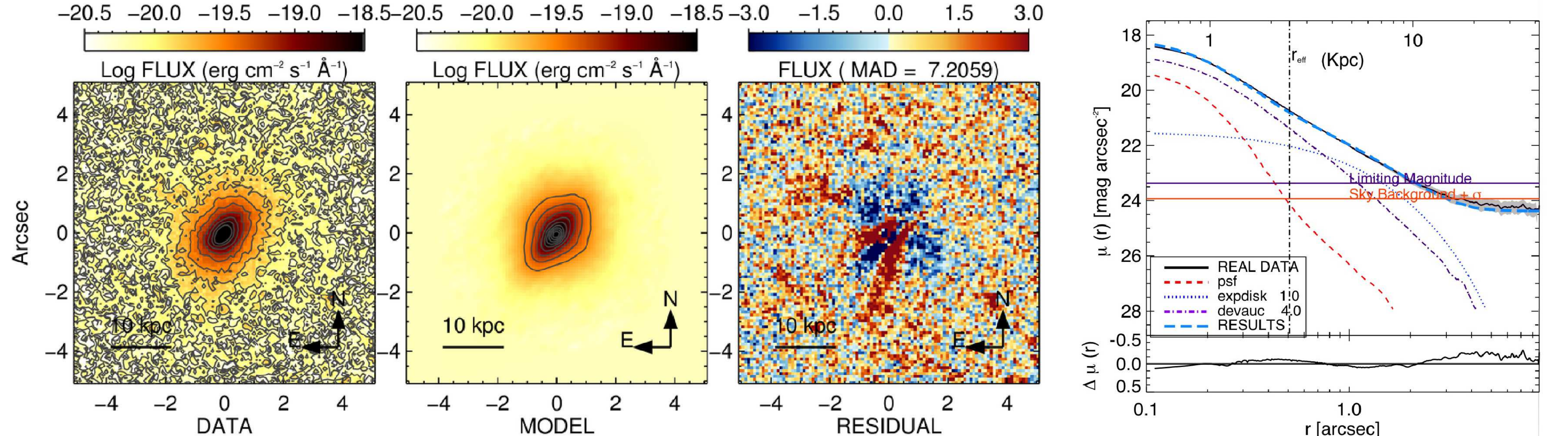} \\
 \vspace{0.1in}
\caption{Examples of {\sc GALFIT} decomposition method. The 2D  images are, from left to right: the HST data, the model  and the residual image. Ten contours are plotted in the two first images, with  values evenly distributed in the range  3$\sigma$ to the maximum flux of the object of interest. The colour scale of the residual images varies within  the range -3$MAD$ and +3$MAD$ (see Sect. 3.2.5). 
The plots on the right panel show the 1D light profile of both data and model,  and the individual structural components. The following convention has been adopted: data (black solid line),  best fit (light blue long-dashed line, labelled ``RESULTS'' in the plots). In addition, a red dashed line will be always used for  point sources; a blue dotted line for disks ($n=$1.0); a purple dash-dotted line for De Vaucouleur profiles (i.e. fixed $n=$4.0);   blue solid lines are  used for  S\'ersic components of free $n$, independently of this index value. The vertical lines mark the effective radius of the model image. The orange and purple horizontal lines indicate the  background level plus $\sigma$ (the standard deviation of the sky background, both calculated in section \ref{noise}) and the limiting magnitude, respectively, calculated as  $-2.5 \times log(\text{sky background}+\sigma) +Zp+5  \times log(\text{pix scale})$ and $-2.5 \times log(\text{sky background}+3 \times \sigma) +Zp+5  \times log(\text{pix scale})$. The grey shadowed area represents   the data Poisson errors.  Bottom panel, inset:  the residuals of the fit $\Delta \rm MAG$=MAG(DATA)-MAG(FIT) at each radial distance are shown.} 
\label{Gpx:0:A}
\end{figure*}

 \renewcommand{\thefigure}{\arabic{figure} (Cont.)}
 \addtocounter{figure}{-1}
 
 \begin{figure*}
\centering
 Disk-dominated:  SDSS J032029.78+003153.5\\
  \vspace{0.1in}
 \includegraphics[width=\textwidth]{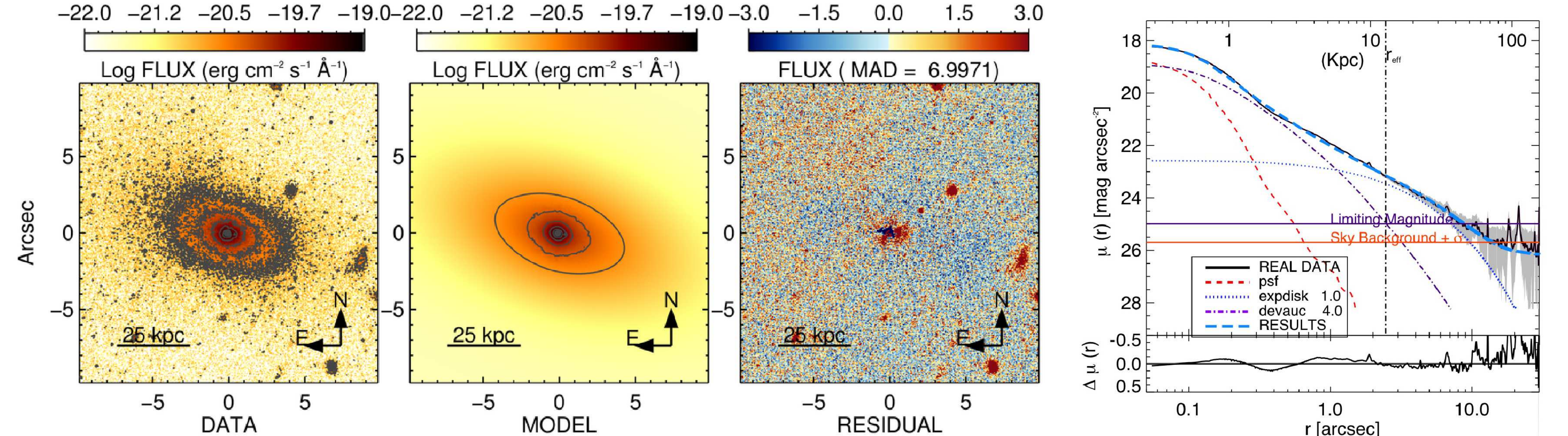} \\
  \vspace{0.1in}
 Bulge-Disk:  SDSS J020234.56-093921.9\\
  \vspace{0.1in}
 \includegraphics[width=\textwidth]{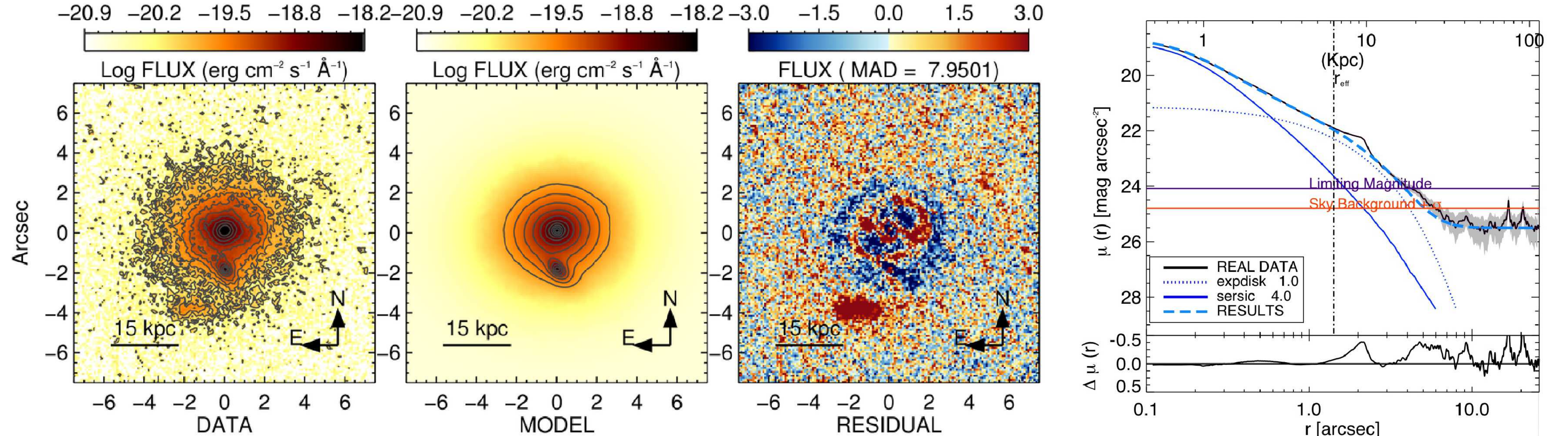}
\vspace{-0.1in}
\caption{} 
\label{Gpx:0:B}
\end{figure*}

\renewcommand{\thefigure}{\arabic{figure}}

 \begin{center}
\begin{table*}
\footnotesize
\begin{tabular}{|l|c|c|c|c|c|}
\hline \hline
 \multicolumn{1}{|c|}{Object }	     &  $lO3$ &   \multicolumn{2}{c|}{Magnitude (mag)} &  Hr {\sc SExtractor} & 	$R_{eff}$ {\sc GALFIT} \\
 \multicolumn{1}{|c|}{SDSS Name } &         & {\sc SExtractor} & {\sc GALFIT}     &(kpc)    & (kpc)		\\
 \multicolumn{1}{|c|}{[1]}		&  [2]  	&  [3]      & [4]         &  [5]   	& [6] 		\\ \hline
  J005515.82-004648.6	&	8.15	&	19.3	&	19.3	&	1.5	&	1.5\\
 J011429.61+000036.7	&	8.66	&	17.7	&	17.5	&	5.5	&	5.9\\
 J011522.19+001518.5	&	8.14	&	18.1	&	18.1	&	7.8	&	4.5\\
 J014237.49+144117.9	&	8.76	&	17.8	&	17.8	&	1.1	&	1.1\\
 J015911.66+143922.5	&	8.56	&	19.7	&	19.5	&	1.4	&	1.2\\
 J020234.56-093921.9	&	8.39	&	17.4	&	17.6	&	7.1	&	6.3\\
 J021059.66-011145.5	&	8.10	&	19.0	&	19.0	&	3.8	&	3.0\\
 J023411.77-074538.4	&	8.77	&	18.7	&	18.7	&	2.2	&	2.3\\
 J031946.03-001629.1	&	8.06	&	18.9	&	18.8	&	6.3	&	7.5\\
 J031927.22+000014.5	&	8.24	&	19.1	&	18.8	&	4.2	&	4.5\\
 J032029.78+003153.5	&	8.52	&	17.3	&	17.0	&	10.0 	&	12.8\\
 J034215.08+001010.6	&	9.08	&	19.3	&	19.2	&	1.2	&	1.1\\
 J040152.38-053228.7	&	8.96	&	18.9	&	19.0	&	2.8	&	2.7\\
 J074811.44+395238.0	&	8.19	&	18.9	&	18.7	&	4.5	&	4.6\\
 J081125.81+073235.3	&	8.88	&	19.1	&	19.2	&	5.5	&	4.0\\
 J082449.27+370355.7	&	8.28	&	19.2	&	19.0	&	1.2	&	1.3\\
 J082527.50+202543.4	&	8.88	&	19.7	&	19.7	&	0.9	&	1.1\\
 J083028.14+202015.7	&	8.91	&	18.6	&	18.5	&	2.2	&	2.5\\
 J084041.08+383819.8	&	8.47	&	17.9	&	17.8	&	6.3	&	7.1\\
 J084309.86+294404.7	&	9.34	&	18.8	&	18.7	&	3.9	&	3.1\\
 J084856.58+013647.8	&	8.46	&	17.8	&	17.1	&	6.1	&	4.9\\
 J084943.82+015058.2	&	8.06	&	19.5	&	19.5	&	1.9	&	2.2\\
 J090414.10-002144.9	&	8.93	&	17.9	&	17.8	&	2.8	&	2.9\\
 J092318.06+010144.8	&	8.94	&	18.9	&	18.8	&	3.6	&	3.1\\
 J094209.00+570019.7	&	8.31	&	18.9	&	18.7	&	1.8	&	1.6\\
 J094350.92+610255.9	&	8.46	&	19.5	&	19.6	&	1.0	&	1.0\\
 J095629.06+573508.9	&	8.38	&	19.1	&	19.0	&	2.8	&	3.2\\
 J100329.86+511630.7	&	8.11	&	18.4	&	18.4	&	5.2	&	4.8\\
 J103639.39+640924.7	&	8.42	&	18.1	&	17.8	&	7.5	&	7.3\\
 J112907.09+575605.4	&	9.38	&	19.3	&	19.1	&	2.0	&	2.4\\
 J113710.78+573158.7	&	9.61	&	19.1	&	18.6	&	3.0	&	3.3\\
 J140740.06+021748.3	&	8.90	&	19.3	&	19.2	&	1.4	&	1.5\\
 J150117.96+545518.3	&	9.06	&	17.2	&	17.2	&	6.2	&	6.1\\
 J154133.19+521200.1	&	8.25	&	17.8	&	18.0	&	7.9	&	6.7\\
 J154613.27-000513.5	&	8.18	&	19.3	&	19.0	&	3.8	&	3.8\\
 J172419.89+551058.8	&	8.00	&	19.5	&	19.3	&	1.5	&	1.4\\
 J172603.09+602115.7	&	8.57	&	19.8	&	19.8	&	1.2	&	1.2\\
 J173938.64+544208.6	&	8.42	&	19.3	&	19.2	&	2.1	&	2.6\\
 J214415.61+125503.0	&	8.14	&	18.4	&	18.5	&	1.9	&	1.2\\
 J223959.04+005138.3	&	8.15	&	18.9	&	18.9	&	3.7	&	3.6\\
 J231755.35+145349.4	&	8.10	&	18.4	&	18.3	&	6.7	&	6.2\\
 J231845.12D-002951.4	&	8.00	&	18.6	&	18.4	&	5.6	&	4.3\\ \hline
    \end{tabular}
\caption{Comparison between the total magnitude and the half-radius calculated with {\sc SExtractor} and the effective radius inferred with the task {\sc ELLIPSE} applied to the  {\sc GALFIT} 2D model image. Only objects that could be fitted with {\sc GALFIT} are shown. The {\sc SExtractor} values were obtained by applying the code to the HST data.  The  {\sc GALFIT} $mag$ and $R_ {\rm eff}$  values were measured in the best fit model image. }
\label{Table:1}
\end{table*}
\end{center}

\subsubsection{PSF} 
 
We refer the reader to the Appendix \ref{PSF}  for a detailed description and discussion on the PSF construction method. 

\subsubsection{Determination of the sky background}  
\label{noise}

An accurate determination of the sky background is essential, especially  for faint objects and galaxies with extended low surface brightness  structures.

Following  \citet{Haussler2007} we introduced the background level as a fixed parameter of {\sc GALFIT}. To calculate it for each AGN image,  we selected  emission  free areas (i.e. masking sources)  around the object to avoid contamination. The sky background was then estimated as the average value of all pixels with values $<3\sigma$, after applying a  3$\sigma$ clipping method.

\subsubsection{Methodology}
\label{Methodology}

  We have followed two steps to obtain the parametric fits:
   
\begin{itemize}

\item    Method Par-I. The light profile and global morphology are parametrized using a single 
S\'ersic component. This method has been used in numerous works (\citealt{Weinzirl2009}, \citealt{Buitrago2013}, \citealt{Davari2016}) to classify galaxies depending on $n$ into  spheroidal or disk galaxies.

 As we will see (Sect. \ref{Re:Bulge-Disc:Class}), single S\'ersic profiles do not provide acceptable fits in the majority of our objects. This method has been useful, on the other hand,   to obtain a preliminary guess of the galaxy types and to  constrain the input parameters of {\sc GALFIT} when applying more complex fits (Method Par-II).  It has also proved useful to identify objects where the contribution of a point source is necessary. It is found that whenever a single S\'ersic component fit  resulted on a  $n\ge$8, the contribution of a point source is confirmed by the more sophisticated Method Par-II. The point source  contribution may be relatively low  in flux, but it can modify the shape of the inner light profile significantly.

\item Method Par-II. Two or more components  are considered in the fit with the goal of isolating and parametrizing different constituents  and, when necessary, to separate overlapping galaxies. The combination can include two or more S\'ersic components and, if necessary, a point source. All are combined to reproduce the global light profile of a given galaxy.

For a given object, we select the fit with the minimum number of components that best reproduces the surface brightness profile and leaves minimum residuals in the 2D  residual image, excluding asymmetric peculiar features.   We find that all  galaxies can be successfully fitted with   a maximum of three components: a point source and two S\'ersic profiles.

\end{itemize}

\subsubsection{Physical nature of the structural components}
\label{Bulge-Disc:Class}

One of our main aims is to classify the galaxies according to the structural component that dominates the galaxy luminosity (i.e. bulge-dominated, disk-dominated, other).  It is necessary to define some criteria to associate each  component  with a physical counterpart.

We assume $n=$1 for disks.  This is a  common practice which is justified by the fact that the light profile of disks is indeed exponential   (\citealt{Freeman1970}). The situation is less simple for bulges and elliptical galaxies. Although the $r^{1/4}$ de Vaucouleurs profile  is often assumed, numerous works have shown that this must be generalized to  $r^{1/n}$ to account for the range of values of $n$ spanned by different galaxies \citep{Trujillo2001a, Haussler2007, Allen2006,  Ribeiro2016}. Following \citet{Gadotti2009} (see also  Barentine \& Kormendy \citeyear{Barentine2012}) we will consider that a  $n\ge$2 S\'ersic is a bulge. 

Peculiar   features are frequent around high-luminosity AGN (e.g. tidal tails, fans, etc). This is also the case for our sample (\citealt{VillarM2012}; see   also \ref{Visual:classification:2}). When such features are irregular and asymmetric, they are easily recognized in the residual images. However, when they are diffuse and symmetric,  {\sc GALFIT}  may  reproduce them successfully with low $n\le$1 S\'ersic components, which  may be erroneously interpreted as disks or bars ($n\sim$0.5, \citealt{Peng2002}). To avoid such degeneracy (which we find to affect a minority of objects anyway),   we have carefully checked for every target whether the interpretation of the nature of the different structural components revealed by the fit is  coherent with the visual inspection.

Taking into account all these considerations,  the galaxies will be classified as follows:

\begin{itemize}

\item Highly-disturbed systems. This group contains objects with strongly distorted morphologies with clear signs of galactic interactions, according to the classification described in Sect. \ref{Visual:class}. These objects cannot be fitted with {\sc GALFIT}  accurately.

\item Point-source-dominated objects. A point source contributes $>$50\% to the total galaxy flux.

\item Objects with a single S\'ersic component (with or without point source). The objects will be classified in terms of $n$ following \citet{Graham2005}  (see also \citealt{Haussler2007}).

\begin{itemize}

\item $n<$2. Disk-like. These will be considered {\it disk-dominated} systems.

\item $n\ge2$. Spheroidal. These will be considered {\it bulge-dominated} systems.

\end{itemize}

\item Objects with two S\'ersic components (with or without point source).

\begin{itemize}

\item Systems with B/D$<$0.8  will be classified as  {\it disk-dominated}. The disk-component ($n=$1) contributes more than the bulge ($n\ge$2).

\item Systems with B/D$>$1.2 will be classified as   {\it  bulge-dominated}. The bulge-component  contributes more than the disk.

\item Systems with B/D$=$1.0$\pm$0.2, will be classified  as  {\it bulge+disk} systems. The disk and the bulge have similar contribution.

\end{itemize}

\end{itemize}

\subsubsection{Selection of the best  fit}  
\label{Selection:best:fit}
 
  Our final parametric galaxy classification  is based on Method Par-II (see Section \ref{Methodology} and Tables \ref{Table:param} and \ref{Table:class}).   
  We explain next the criteria adopted to  select the best fit for each object. 

Five examples are shown in Fig. \ref{Gpx:0:A}. For each object the HST HLA image, the 2D  {\sc GALFIT}  model  and the residual image are shown on the left panels.
 The  surface brightness profile, the best fit, the profiles of the individual structural components and the residuals (data - model) are plotted in the right panel.

The residual 2D image has been created as follows. The best fit model was subtracted from the original image producing a residual image $res$. This was then divided by MAD (the median absolute deviation) calculated using  regions with no bright objects or features.  Pixels with absolute values $|F|=|\frac{res}{MAD}|<$3 keep their values and they will be considered to be within the noise level. Pixels with  $|F|\ge$3  are replaced by +3 or -3 depending on the $F$ sign. These residuals are potentially real (at or above the detection limit), except possibly in the central region of galaxies, since  a small spatial shift of a fraction of a pixel of the PSF relative to the galaxy centre can produce strong artificial residuals. The colour scale in the final residual image presented in the figures is in the range
-3$\times MAD$  to +3$\times MAD$. 

    Multiple combinations of structural components were attempted based on different assumptions: from the standard  $n = 1$ (disks) and $n = 4$ (bulges), to combinations of free $n$ components, including or not a point source.

The selection of the best fit for each object  is based on the following criteria/checks:
    
   \begin{itemize}
   
\item The fit consists of the minimum number of components that best reproduces the surface brightness profile and leaves minimum residuals in the 2D  residual image, excluding peculiar features that are confirmed  to be real.

\item  Fits requiring S\'ersic components whose contribution to the total flux is $<$10\% ($>$2.5 magnitudes difference  relative to the total magnitude) are rejected. It is found that these   components have always low $n$ and do not result in significant changes in the fit.  This does not apply to the point source: since its profile is very steep (high $n$), even  a small contribution of $\sim$a few \% can change significantly the shape of the central regions of the galaxies.
  
\item   All fits were visually inspected and compared with the original image. This was sometimes very useful, for instance, to confirm the presence of disks; also to check  whether low $n$ components are bars/disks or, alternatively, peculiar diffuse features (Sect. \ref{Bulge-Disc:Class}).

\item An additional test that helped to refine (even discard) some fits was applied as follows. A region centred on the galaxy centroid and of  typical size $\sim$25 pixels $\times$ 25 pixels (2.5$\times$2.5 arcsec$^2$ for the WFPC2 and 1.25$\times$1.25  arcsec$^2$  for the ACS) is selected both in the original and  the residual images. The  pixel with maximum  value, $F_{\rm max}$ is identified in he HST image. We  then calculate $\Delta_{\rm max}=\frac{F_{\rm max}-F'_{\rm max}}{F_{\rm max}}$, where $F'_{\rm max}$ is the flux value of that same pixel in the residual image and $\Delta_{\rm min}=\frac{F_{\rm max}+F'_{\rm min}}{F_{\rm max}}$, where $F'_{\rm min}$ is the minimum value measured in the same area in the residual image. When $\Delta_{\rm max}$ or $\Delta_{\rm min}$ are larger than 0.5, the fits are further inspected. Such large deviations, warn about possible problems with the centring, the possible need for a point source or the presence of   peculiar features such as prominent dust lanes. These objects were analyzed with special care to identify whether  the strong residuals are real or artefacts.

\end{itemize}

\section{Results}
\label{Results}

\subsection{Morphological Visual classification}
\label{Visual:conclusion}

\subsubsection{Method Vis-I}
\label{Visual:classification:1}

\begin{figure}
\includegraphics[width=\columnwidth]{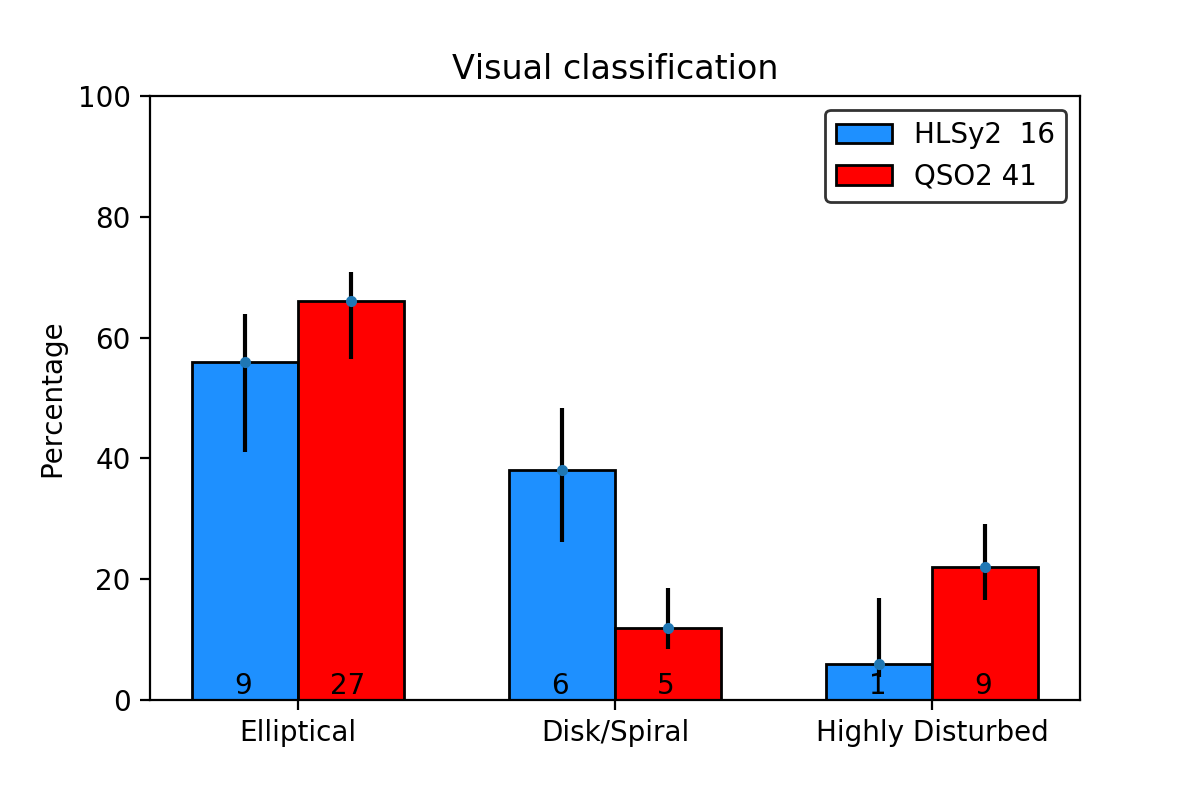}
\caption{Results of the  visual classification Method Vis-I (Sect. \ref{Visual:class}).  The numbers within the bins in this and other histograms  are the actual number of objects classified within each specific bin. The error bars  in this an all histograms are 1$\sigma$ Dirichlet multinomial distribution confidence intervals.}             
\label{Hist:1}
\includegraphics[width=\columnwidth]{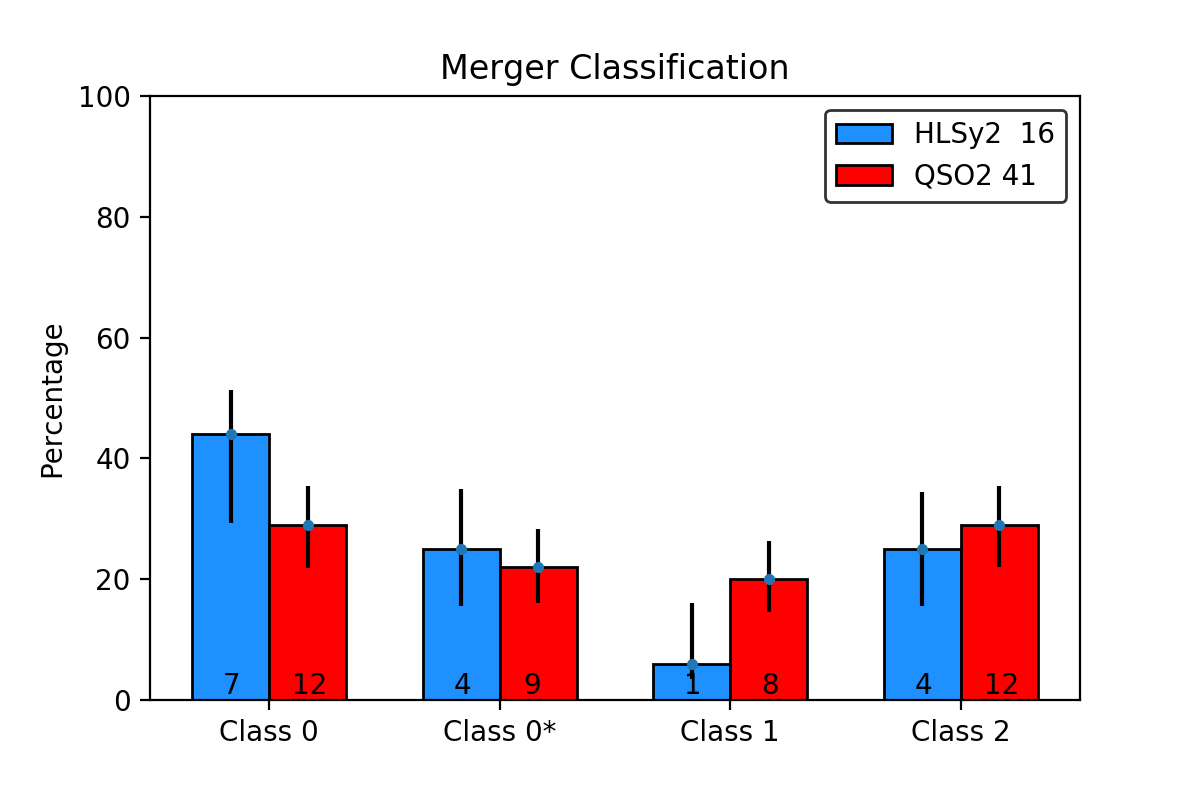}
\caption{Results of the  visual classification Method Vis-II (Sect. \ref{Visual:class}), which is focused on the evidence of merger/interactions. Numbers within the bins and  error bars  as in Fig. \ref{Hist:1}.}
\label{Hist:0}
\end{figure}

We show in  Table \ref{Table:class}  (column [3])  and Fig. \ref{Hist:1} the results of the visual classification based on Method Vis-I described in Sect. \ref{Visual:class}.  Owing to the small sample size,  we estimate the 1$\sigma$ confidence intervals of the different galaxy populations studied here following a similar method as in \citet{Cameron2011}. In our case we use a Dirichlet distribution, a multivariate generalization the Beta binomial distribution, which provides a better performance at low sampling conditions compared to other methods such as the `normal approximation' and the \citet{Clopper1934} approach (see \citealt{Cameron2011} for more details).

The main results are:

\begin{itemize}

\item Among QSO2, 27/41 or 66\%$^{+5}_{-10}$ are visually classified as ellipticals, 5/41 or 12\%$^{+7}_{-4}$  are spirals or disks and 9/41 or 22\%$^{+7}_{-6}$ are highly-disturbed systems (HD). 
 Among HLSy2, 9/16 or  56\%$^{+9}_{-15}$ are ellipticals, 6/16  or 38\%$^{+10}_{-12}$ are  spirals or disks and 1/16 or 6\%$^{+11}_{-3}$ is a HD.

\item Thus, a minority of QSO2  are hosted by disk/spirals. This fraction is significantly higher in the HLSy2  subsample. 
  
 \item There is tentative evidence for the fraction of HD  to be higher in  QSO2  than in HLSy2, although  taking  uncertainties into account the difference is not significant.

 \end{itemize}

\begin{figure*}
\centering
\includegraphics[width=\columnwidth]{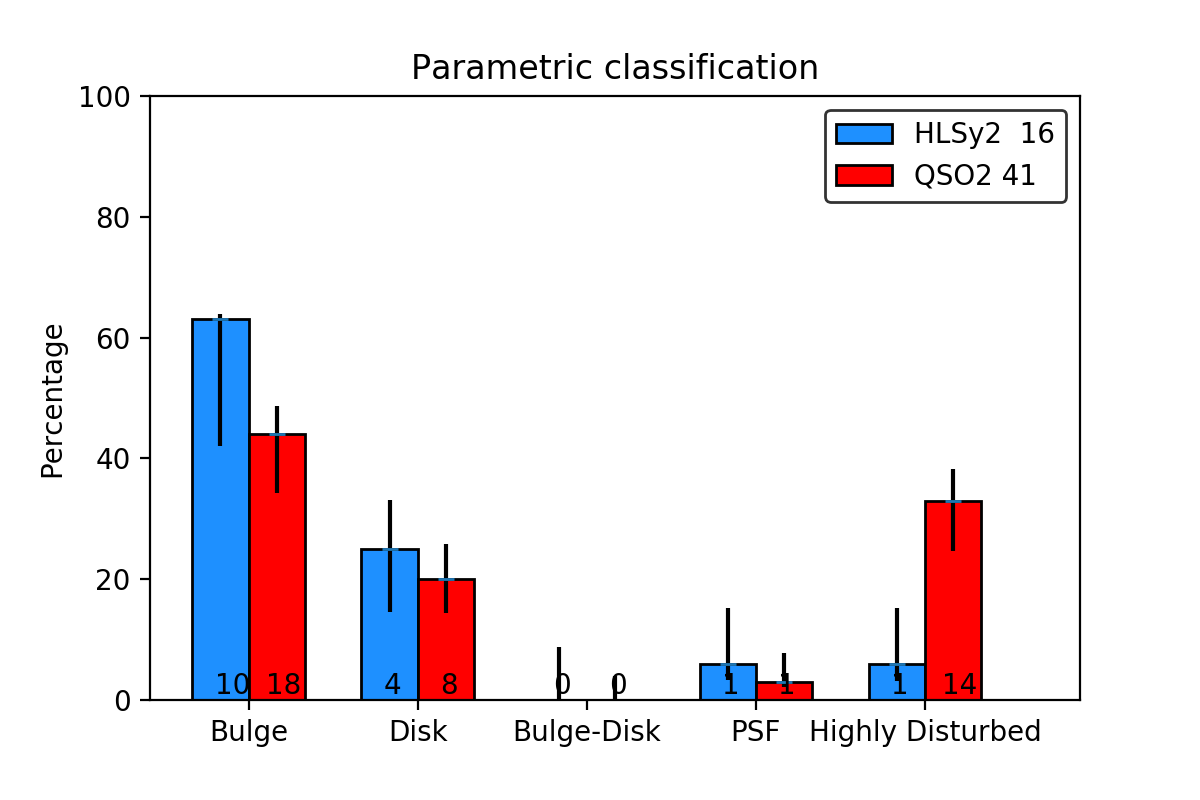}
\includegraphics[width=\columnwidth]{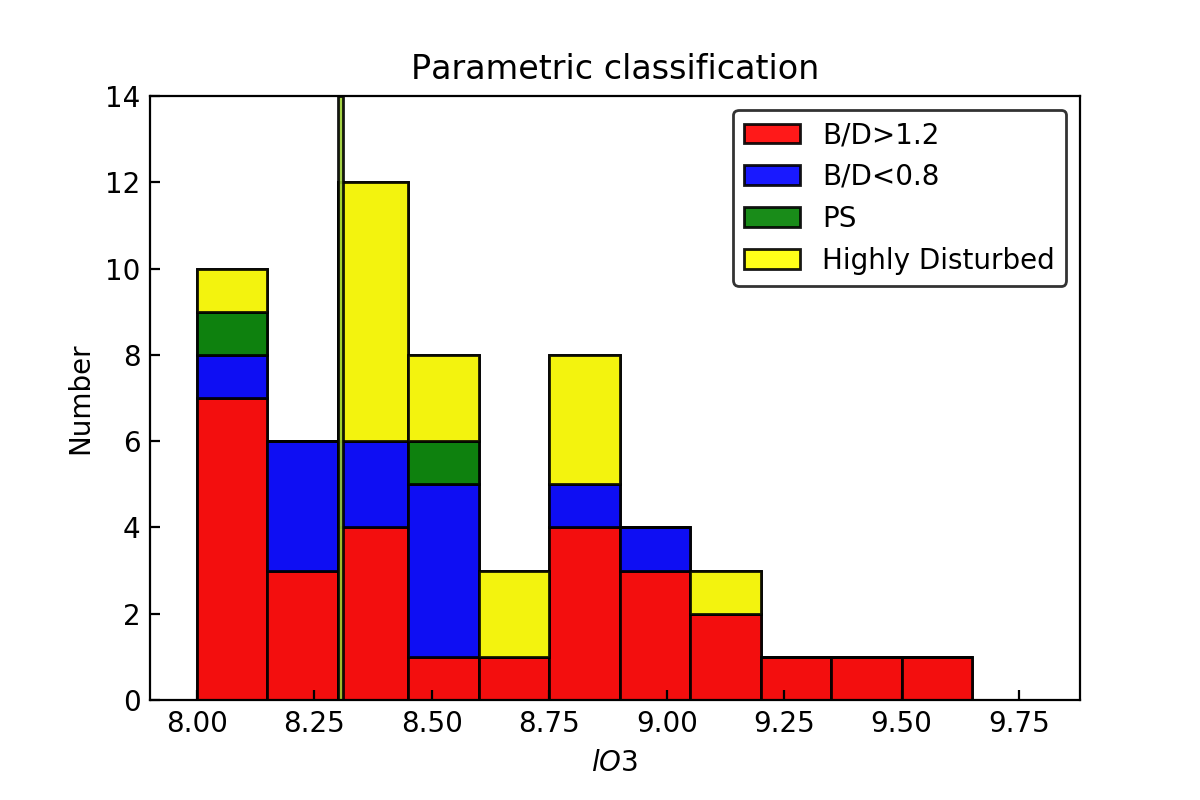}
\caption{{\it Left:} Results of the parametric classification for QSO2 and HLSy2. ``Bulge" are bulge-dominated systems, which include galaxies with B/D$>$1.2 and spheroidal systems (single S\'ersic with $n\ge$2). ``Disk" are disk-dominated systems,  which include galaxies with B/D$<$0.8 and disk like systems  (single S\'ersic with $n<$2). ``Bulge-Disk'' are systems for which 0.8$\le$B/D$\le$1.2 (zero found in the sample) and ``PSF" are  systems where a point souce  contributes $\ge$50\% of the total light. The numbers within the bins and the error bars are explained in Fig. \ref{Hist:1} 
{\it Right:}  Dependence of the galaxy classification with  $L_ {\rm [OIII]}$ (proxy of AGN power). For each luminosity bin, the number of galaxies classified within a certain class is indicated by the height of the corresponding coloured rectangle. For instance, in the the $lO3\sim$8.25 bin there  are 3 bulge-dominated and 3 disk-dominated galaxies. The vertical line corresponds to $lO3$=8.3, assumed  as the dividing value between HLSy2 and QSO2. }             
\label{Hist:2}
\end{figure*}


\subsubsection{Methods Vis-II and Vis-III}
\label{Visual:classification:2}

 The previous visual method does not allow the identification of all objects with signs of mergers/interactions, but only the most morphologically disturbed systems. However, their identification can be done with the classification method Vis-II (see Sect.  \ref{Visual:class}).  The results are shown in Fig. \ref{Hist:0}  (see also Table \ref{Table:class}, column [4]).

The first result is that  QSO2 and HLSy2 are distributed  quite evenly among the 4 classes, from  isolated objects, without (Class 0) or with  (Class 0$^*$) peculiar features,   to objects with signatures of galaxy interactions at different stages (Classes 1 and 2). This means that powerful nuclear activity occurs both in isolated objects and at different  phases of galactic interactions, as already found in different studies such as  \citet{CRamos2011} and \citet{Bessiere2012}.

The additional classification of the peculiar features based on \citet{CRamos2011} (Method Vis-III) provides complementary information: $71\%^{+6}_{-8}$ of QSO2 show peculiar features that have been classified according to their aspect in column [5] of Table \ref{Table:class}.  For the HLSy2 sub-sample the percentage is  $56\%^{+12}_{-13}$. These values may be   lower limits, given the higher difficulty to identify peculiar features in galaxies with spiral/disk structures.

 \citet{Bessiere2012} studied and classified the peculiar features in a complete sample of 20 SDSS QSO2 at 0.30$< z <$0.41 and with $lO3\ge$8.5. 13 of these are also in our sample. They used  deep Gemini Multi-Object Spectrograph-South optical broad-band images obtained with the $r'$-band filter
($r_{\rm G0326}$, $\lambda_{\rm eff}$ = 6300\AA, $\Delta\lambda$=1360 \AA). They found that $\sim$75\% of their QSO2 show evidence for  peculiar features. If we focus on those objects in the HST sample with $lO3\ge$8.5,  we find the same rate as them:  $75\%^{+7}_{-11}$  show peculiar features.

Considering the   13 QSO2 that overlap with our study, \citet{Bessiere2012}  confirm peculiar features in 11 objects. We confirm them in 10. The  discrepant object   is SDSS J011429.61+000036.7. They identify a second nucleus (which, given the unknown $z$, we have classified as projected companion ``PC'') and a shell which is not clearly detected in the HST image.

\subsection{Parametric classification}
\label{Re:Bulge-Disc:Class}

 Following Sect.  \ref{Parametric:class}, we classify the host galaxies of our sample   based on the dominant structural component identified as a result of  the parametric fits. This method could not be applied to 15/57 objects, of which 14 are QSO2 and 1 is HLSy2. In general, they present strongly distorted morphologies. These objects will be referred to as ``highly-disturbed" (HD), in coherence with the visual classification.  The results of the parametric method for individual objects are shown in  Table  \ref{Table:class}. The  distribution of the sample among the different classification groups  is shown in Tables  \ref{compa-merger} and \ref{compa}, and in Fig.  \ref{Hist:2} (left).  The difference between these two tables is that Table   \ref{compa-merger} includes HD systems, while Table  \ref{compa}  does not.


The main results are:

\begin{itemize}

\item   Bulge-dominated systems: This group includes spheroidal galaxies or galaxies with B/D$>1.2$ (Sect. \ref{Bulge-Disc:Class}). It is the most numerous group both for  QSO2 (18/41 or 44$\%^{+5}_{-10}$) and for HLSy2  (10/16 or 63$\%^{+1}_{-21}$).  Taking uncertainties into account, no significant difference between  QSO2 and HLSy2  is found. 

If HD are excluded, no significant difference is found either (18/27 or 67\%$^{+4}_{-15}$ for QSO2 and HLSy2 (10/15 or 67\%$^{+3}_{-21}$).

\vspace{3mm}
 \item   Disk-dominated systems:  These are disk-like galaxies or galaxies with B/D$<0.8$ (Sect. \ref{Bulge-Disc:Class}). Only 25\%$^{+8}_{-11} $(4/16) of HLSy2 and 20\%$^{+6}_{-6}$ (7/41) of QSO2 galaxies are disk-dominated. The difference between both groups is not significant. 

If  HD systems are excluded the fractions become  4/15 or 27\%$^{+10}_{-11}$ for HLSy2  and 8/27 or $\sim$29\%$^{+8}_{-9}$ for QSO2. 
\vspace{3mm}
\item  Bulge+disk systems  (B/D=1.0$\pm$0.2) have not been found  (0\%$^{+9}$ of HLSy2 and  0\%$^{+3}$ of QSO2).  
\vspace{3mm}
\item Disks (not necessarily dominating the total galaxy flux) are identified in a significantly higher fraction of HLSy2 (7/16 or 44$\%^{+13}_{-11}$) than QSO2 (10/41 or 24$\%^{+8}_{-6}$).

If   HD systems are excluded, the difference between both fractions disappears:  HLSy2 (47\%$^{+13}_{-12}$) and QSO2 (37\%$^{+10}_{-8}$) have disks.
\vspace{3mm}
\item   A point source component is isolated in a high fraction of objects: 29 (10 HLSy2 and 19 QSO2) of the 42 (69\%$^{+7}_{-8}$) for which the parametric analysis could be applied, with no significant difference between both groups. The relative contribution to the total flux varies between 3\% and 51\%, with  average value 20.7$\pm$2.9\% (median 14.2\%). The PSF dominates ($\ge$50\% of the total flux) in just one HLSy2 and one QSO2.

\end{itemize}

\subsection{Dependence of galaxy host with $lO3$}
\label{bt}

We have seen that, excluding highly-disturbed systems, the parametric classification of the HLSy2 and QSO2 hosts are consistent within the errors. 

 We perform next a more  detailed analysis of the dependence of galaxy properties  with  $lO3$,  proxy for AGN power. For this, we use a finer sampling of the line luminosity range, instead of the coarse and somewhat arbitrary division in  HLSy2 and QSO2 at  threshold  $lO3$=8.3.  The results are shown in  Fig. \ref{Hist:2} (right).

A  clear dependence of the galaxy properties on  AGN power is revealed. While bulge-dominated systems spread across the total range of $lO3$, disk-dominated galaxies concentrate mostly at   $lO3$$\la$8.6. This is in fact closer to the dividing luminosity  $lO3$=8.5  between QSO2 and HLSy2 assumed by \citet{Zakamska2003} than to the 8.3 value assumed by \citet{Reyes2008}.
Considering the full sample, there are 10/36 or 28\%$^{+8}_{-6}$ disk-dominated galaxies below  $lO3$=8.6 and  2/21 or 10\%$^{+10}_{-3}$ above.

The differentiation is even clearer when highly-disturbed systems are excluded. 38\%$^{+10}_{-8}$  objects with  $lO3$$<$8.6 are disk-dominated    versus   13\%$^{+14}_{-5}$  above this luminosity.   There are  56\%$^{+9}_{-10}$  bulge-dominated galaxies at $lO3$$<$8.6   and 87\%$^{+5}_{-14}$ at $lO3$$>$8.6.

\begin{figure}
\centering
\includegraphics[width=\columnwidth]{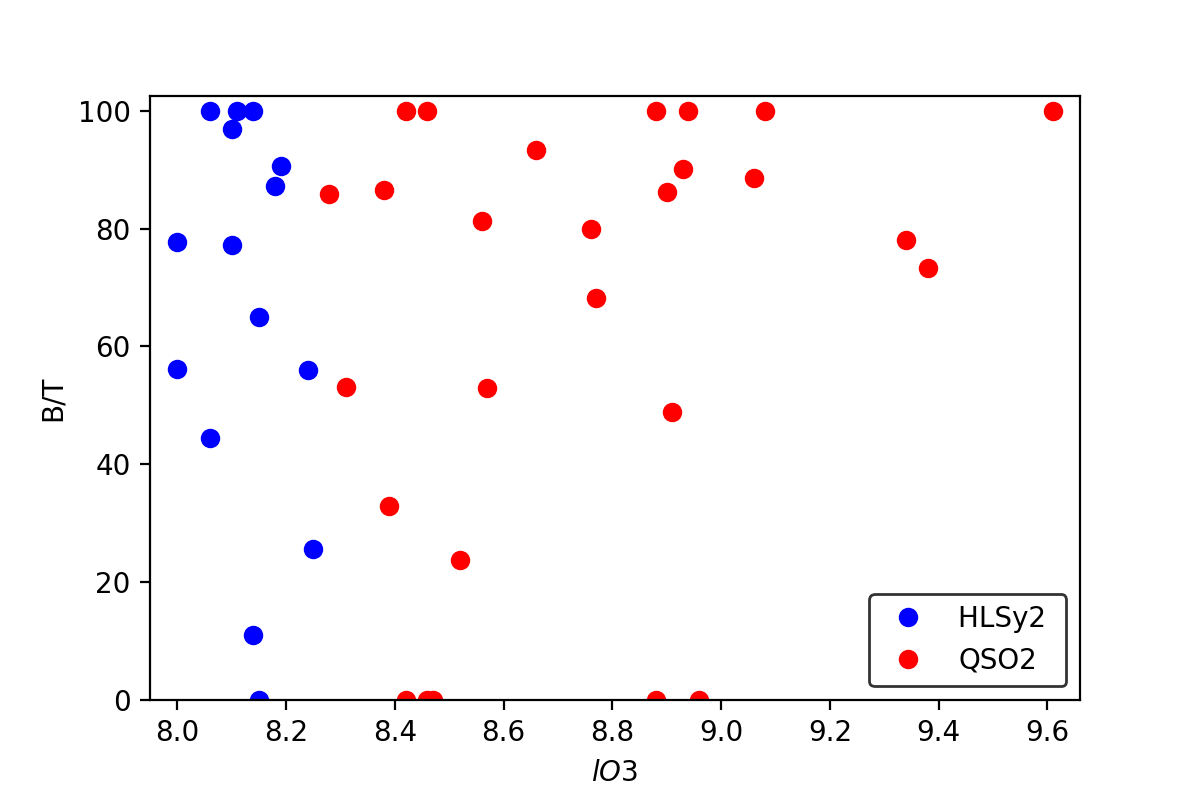}
\caption{Relative contribution of the spheroidal-component to the total galaxy light (B/T)  vs. $lO3$. Only  galaxies that could be fitted with {\sc GALFIT} are plotted. B/T increases with AGN power. For $lO3\ga$8.6, most galaxies are bulge-dominated.}             
\label{dunlop}
\end{figure}

The increasing incidence of bulge-dominated systems with AGN luminosity is also apparent when we study the variation with   $lO3$   of the  relative contribution of the spheroidal-component to the total galaxy light (B/T)  for  the objects that could be fitted with {\sc GALFIT} (Fig. \ref{dunlop}).   The average B/T increases with AGN power. Most objects  with  $lO3$$\ga$8.6 have B/T$\ga$70\%, while  at lower luminosities, the galaxies span the full range of possible B/T values.

\subsection{Contribution from a point source}
\label{psf}

\begin{table}
\small
\begin{tabular}{lccc}
\hline \hline
Parametric Class             	& AGN type 	& With PS 		& Without PS 	\\ \hline
\multirow{2}{*}{Bulge-dominated} 	& HLSy2      	&  (6/10)  60\%$^{+4}_{-24}$  	& (4/5) 80\%$^{+1}_{-35}$		\\
 						  	& QSO2    	&  (11/19) 58\%$^{+5}_{-17}$	& (7/8) 88\%$^{+1}_{-28}$		\\
\multirow{2}{*}{Disk-dominated}		& HLSy2		&  (3/10)  30\%$^{+10}_{-14}$	& (1/5) 20\%$^{+20}_{-10}$	\\
							& QSO2		&  (7/19)  37\%$^{+12}_{-8}$	& (1/8) 13\%$^{+17}_{-6}$		\\
\multirow{2}{*}{PSF-dominated}		& HLSy2		&  (1/10)  10\%$^{+14}_{-5}$	& N/A	\\
							& QSO2		&  (1/19) 5\%$^{+9}_{-2}$ 	& N/A		\\  \hline
 \end{tabular}
\caption{Comparison between the galaxy classification of HLSy2 and QSO2 with and without a point source. The fractions are quoted in brackets. Tentative evidence is hinted for a higher fraction of bulge-dominated systems in objects without a point source.}
\label{Table:contpointsource}
\end{table}

A  point source has been isolated in 29  of the 42 objects  (69\%$^{+7}_{-8}$) for which the parametric method could be applied. The relative contribution to the total light of the galaxy in this subsample is in the range Light Fraction ($L.F.$) $\sim$3-51\% with median value 14.2\% and standard deviation 14.8\%. Even when $L.F.$ is small ($\sim$few\%), this cannot be ignored in the fits, since the structural parameters of the hosts can be severely affected.

Our results are in good agreement with  \citet{Inskip2010}  (see Appendix \ref{otherworks} for a description of their sample). They  found that the K-band  images of 17 narrow line radio galaxies (NLRG) are often contaminated by  a point source. They identified this component  in  12 objects with $L.F.$ in the range $\sim$1-36\%,  and with median and standard deviation values 11.0\% and 10.9\% respectively.

This unresolved component is  a combination of different sources whose relative contribution changes with spectral range. While in  \citet{Inskip2010} an enhanced  contribution of the AGN direct light  may play a role  due to  less severe extinction effects in the near infrared, in our data the contamination by strong emission lines emitted by the compact narrow line region (NLR) is possibly high in many objects (see Table 1). Compact continuum sources are also potential contributors such as nebular continuum associated with the  NLR, scattered AGN light  and nuclear starbursts (\citealt{Dickson1995}, \citealt{Bruce2015}, \citealt{Bessiere2017}).

We have compared the galaxy host classification for objects with and without point source (Table \ref{Table:contpointsource}).
The statistics is  very poor and the errors  large, so that significant differences cannot  be claimed.  On the other hand,  tentative evidence is hinted  for a higher fraction of bulge-dominated systems among objects (both QSO2 and HLSy2) with a point source.  

We cannot discard that this is an artificial effect. We may be missing a point source in some objects, where this contibution cannot be clearly recognized by our fitting method.   Ignoring the point souce  would result on steeper S\'ersic profiles (and, thus, higher $n$ values) for the central region of the galaxy hosts so that  some may be miss-classified as bulge-dominated as a consequence.  If this is the case, the fraction of bulge-dominated systems in galaxies with no point source would represent an upper limit.

The other alternative is that there is a real intrinsic difference between both groups (with and without point source contribution). Understanding the physical origin of such difference, if confirmed, would be of great interest possibly pointing to differences in openning angle of the central obscuring strucuture and/or the amount of obscuring dust in the central regions and/or orientation.

 \subsection{Kormendy relation}
\label{kormendy}

\begin{figure}
\centering
\includegraphics[width=\columnwidth]{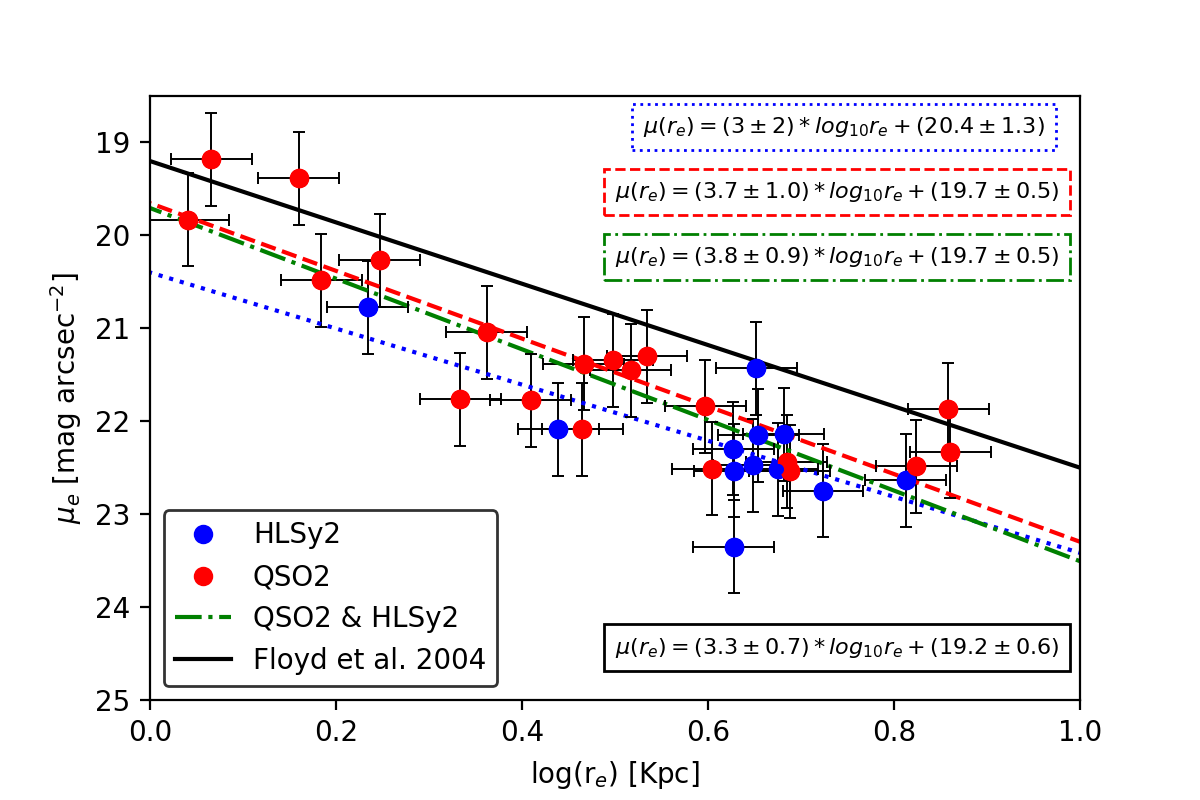}
\caption{Relation between surface magnitude and effective radius for the spheroidal galaxies and bulges in our sample of luminous AGN. The best fits are shown for the QSO2, the HLSy2 and both groups together. The slope is consistent with that found by 
Floyd et al. (2004) for the bulge-dominated hosts  of a sample of 14 QSO1 at $z\sim$0.4.}             
\label{kormendy_fig}
\end{figure}

 We show in Fig. \ref{kormendy_fig}  the Kormendy relation for our sample of host galaxies (\citealt{Hamabe1987}). The effective radius,  $r_{\rm e}$, and the surface brightness magnitude   at  $r_e$,  $\mu(r_e)$, are plotted only for the spheroidal component of galaxies. Errors on $r_{\rm e}$ are expected to be typically $\sim$10\% \citep{Buitrago2018}. We have assumed $\Delta\mu_{\rm e}$=0.5 magnitude errors on    $\mu(r_e)$.   This is based on the study of how  $\mu_{\rm e}$  varies  for a 10\% change in $r_e$ using different S\'ersic profiles.  This variation, $\Delta\mu_{\rm e}$, is clearly larger in steeper (high $n$) profiles  and it can be up to $\sim$0.5 mag.  This is the error we have used  which is therefore conservative. 

 The following relations are found:
$${\rm HLSy2:} ~\mu(r_{\rm e}) = (3.0\pm2.0) \times log(r_{\rm e}) + (20.4 \pm 1.3)$$
$${\rm QSO2:} ~\mu(r_{\rm e}) = (3.7\pm1.0) \times log(r_{\rm e}) + (19.7 \pm 0.5)$$
$${\rm HLSy2~ + ~ QSO2:} \mu(r_{\rm e}) = (3.8\pm0.9) \times log(r_{\rm e}) + (19.7 \pm 0.5)$$

The three trends have correlation coefficients $\rho^2=$0.46, 0.76 and 0.71 respectively (the weak trend for HLSy2 is biased by the outlying measurement of the more compact system with log(re)$\sim$0.2 kpc).

These relations are consistent within the errors with the  scaling relation found by \citet{Bernardi2003} for 9000 SDSS early-type galaxies  at $0.01\le z\le$0.3 (slope 3.33$\pm$0.09).

It is also consistent with the relation found for the bulge-dominated host galaxies  in   \citet{Floyd2004} sample of QSO1 at $z\sim$0.4 (Appendix \ref{otherworks} for a description of their sample; see also \citealt{Dunlop2003}). 
 
$$\mu(r_{\rm e}) = (3.3\pm0.7) \times log(r_{\rm e}) + (19.2 \pm 0.6)$$

 Our sample is shifted down the vertical axis by $\sim$0.6 magnitudes with respect to \citet{Floyd2004} QSO1, although the two samples are within the scatter both have ($\Delta\mu_{\rm e}\sim\pm0.6$).  Both samples are  basically at identical $z$, and thus  redshift dimming does not play a role. We do not believe it is a consequence of the filters used either: 10 of  17 objects  of \citet{Floyd2004}  were observed with the same filter F814W  as the majority of our sample. The other 7 were observed with the F791W filter, redder  and slightly narrower.  Several objects in our sample were observed with the ACS   and with a significantly narrower filter than  F814W. These objects are not shifted in any sense relative to the rest of the objects. The filters, therefore, do not seem to have any influence. Another possible effect is that \citet{Floyd2004} plot $\mu_ {\rm e}$ and $r_{\rm e}$ of the galaxies, and not only the spheroids, as we have done. Although all are  bulge dominated, not removing the contribution of a disk  may result on brighter magnitudes. A final possibility is that the QSO1 are hosted by more luminous and more massive spheroids   (see below).

The average $\langle r_{\rm e} \rangle$ of the {\it spheroidal component}  are  3.4$\pm$0.5 kpc (median 3.0$\pm$1.2 kpc)  and  4.3$\pm$0.4 kpc (median 4.5$\pm$0.3 kpc) for QSO2 and HLSy2 respectively. 
 For comparison \citet{Greene2009} obtain $\langle r_{\rm e} \rangle$=3.1$\pm$0.8 kpc  (median 2.6$\pm$0.5 kpc) for their sample of luminous QSO2 (Appendix \ref{otherworks})\footnote{These values may be somewhat underestimated since the contribution of a point source is not taken into account}. These results tentatively suggest that QSO2 may be hosted by smaller spheroids than HLSy2. The uncertainties  due to the poor statistics are however too large to confirm this.

 We have also measured  $r_{\rm e}$ for the total galaxies (spheroidal + disk components when present),   excluding the point source, which can result on an underestimation of $r_ {\rm e}$ even when the flux contribution is low ($<$10\%)\footnote{For comparison, if we do not take into account the contamination by the point source in our sample,  $\langle r_{\rm e} \rangle$  is 3.5$\pm$0.6 for QSO2 and 4.5$\pm$0.3 kpc for HLSy2 versus $\langle r_{\rm e} \rangle$=3.9$\pm$0.6 kpc and 5.0$\pm$1.5 kpc respectively}.  We obtain 
$\langle r_{\rm e} \rangle$=3.9$\pm$0.6 kpc (median 3.3$\pm$1.6 kpc) for QSO2 and  5.0$\pm$1.5 kpc (median 4.8$\pm$0.9 kpc)  for HLSy2.

The QSO2 sizes are significantly smaller than those measured for QSO1 and NLRG at similar $z$ by other authors\footnote{For this comparison, we have converted  all $r_{\rm e}$ values in other works to the  Cosmology used by us}. \citet{Inskip2010} sample of NLRG have $\langle r_{\rm e} \rangle=$10.1$\pm$1.6 kpc (median 10.0 kpc).   \citet{Dunlop2003}  radio-quiet and radio-loud QSO1 have  $\langle r_{\rm e} \rangle=$7.6$\pm$1.2 kpc  (median 6.4  kpc) and  $\langle r_{\rm e} \rangle$=8.2$\pm$0.8 kpc  (median 8.0 kpc) respectively.  \citet{Floyd2004} obtain $\langle r_{\rm e} \rangle=$7.2$\pm$1.3 kpc  and 6.1$\pm$1.3 kpc for radio-loud and radio-quiet QSO1 respectively. Finally,      $\langle r_{\rm e} \rangle=$7.7$\pm$3.6 kpc for \citet{Falomo2014}  QSO1 sample.

 \citet{Shen2003} found that  early type  galaxies  ($n>$2.5) at $z\sim$0  with  log($M_{\ast}/M_{\rm \odot}$) in the range 10.0-11.0 have $r_{\rm e}\sim$2.1-4.2 kpc.   Only galaxies with log($M_{\rm *})\ga$11.4 have sizes $>$7 kpc. \citet{Wylezalek2016} inferred log($M_{\ast}/M_{\rm \odot}$)=9.8-11.0 for a sample of 20 QSO2 at similar $z$ as our sample, with median and average values 10.4 and 10.5 respectively\footnote{We have scaled their $M_{\ast}$  to a \citet{Kroupa2001} Initial Mass Function for comparison with  \citet{Shen2003}}.   

Therefore,  the $\langle r_{\rm e} \rangle$  we obtain for our QSO2 are in reasonable agreement with those expected based on the typical stellar masses inferred for other luminous QSO2. It is possible that the QSO1 and NLRG samples mentioned above are hosted by more massive galaxies.

\section{DISCUSSION}
\label{Discusion}

Studies of  low  $z$ QSO1 ($z\la$0.5)  since the era of HST show that luminous quasars (radio-loud and radio-quiet) are hosted by a diversity of host galaxies, including ellipticals as bright as the brightest cluster ellipticals, normal ellipticals, spirals and highly-disturbed interacting systems (e.g. \citealt{Bahcall1997}, \citealt{Kukula2001}, \citealt{Dunlop2003}, \citealt{Percival2001}, \citealt{Floyd2004}, \citealt{Falomo2014}).

\begin{landscape}
\begin{table}
\scriptsize
\begin{tabular}{lccccccccccccccccc}
\hline \hline
					&		& \multicolumn{2}{c}{Total} & \multicolumn{2}{c}{Point Source} & \multicolumn{6}{c}{$1st$ Profile Component} 				& \multicolumn{6}{c}{$2nd$ Profile Component} 								 \\
  \cmidrule(l){3 - 4}  \cmidrule(l){5- 6}  \cmidrule(l){7 - 12}  \cmidrule(l){13 - 18} 
\multicolumn{1}{c}{SDSS}&	Method	& Mag   	& $R_{eff}$	& Mag  		& Light 	& Mag       & Light		& $R_{eff}$ & $n$ 	& $b/a$ 		& $PA$ 		& Mag   	& Light      & $R_{eff}$ & $n$    & $b/a$ 		& $PA$ 		 \\
\multicolumn{1}{c}{Name}& {\sc GALFIT}& (mag) 	&  (kpc)    & (mag) 	& Fraction 	& (mag)		& Fraction 	& (kpc)	   	&	 	&	   		&($\textordmasculine$)& (mag)	& Fraction & (kpc) 	&	    &        	&(\textordmasculine)        	  \\
\multicolumn{1}{c}{[1]}	&  [2]      &  [3]  	&  [4]  	&  [5]  	&  [6]		& [7] 		& [8] 		& [9]		& [10] 	& [11]		& [12]    	& [13]      & [14]      & [15]		& [16] 	&  [17]  	& [18]	\\  \hline 
J005515.82-004648.6 &	2B		&	19.3	&	1.5		&	20.4	&	35.00	&	19.7	&	65.00	&	2.4		&	1.2	&	0.57	&	68.5	&	   		&			&			&		&			&		\\
J011429.61+000036.7 &	2B		&	17.5	&	5.9		&	21.3	&	6.70	&	18.2	&	93.30	&	13.0	&	4.0	&	0.74	&	-58.7	&	   		&			&			&		&			&		\\
J011522.19+001518.5 &	1		&	18.1	&	4.5		&	   		&			&	18.1	&	100.00	&	6.3		&	4.0	&	0.59	&	-52.8	&	   		&			&			&		&			&		\\
J014237.49+144117.9 &	2B		&	17.8	&	1.1		&	20.4	&	20.00	&	18.8	&	80.00	&	1.5		&	4.0	&	0.75	&	-52.4	&	   		&			&			&		&			&		\\
J015911.66+143922.5 &	2B		&	19.5	&	1.2		&	22.1	&	18.67	&	20.4	&	81.33	&	1.9		&	1.4	&	0.81	&	-46.4	&	   		&			&			&		&			&		\\
J020234.56-093921.9 &	2A		&	17.6	&	6.3		&	   		&			&	18.7	&	32.97	&	2.4		&	4.0	&	0.81	&	-78.8	&	18.0	&	67.03	&	8.8		&	1.0	&	0.85	&	-81.3	\\
J021059.66-011145.5 &	2B		&	19.0	&	3.0		&	20.7	&	22.80	&	19.2	&	77.20	&	8.2		&	2.5	&	0.48	&	69.9	&	   		&			&			&		&			&		\\
J023411.77-074538.4 &	2D		&	18.7	&	2.3		&	22.0	&	10.38	&	19.8	&	68.29	&	2.5		&	2.8	&	0.86	&	81.5	&	21.2	&	21.32	&	4.2		&	0.3	&	0.85	&	64.0	\\
J031946.03-001629.1 &	2D		&	18.8	&	7.5		&	   		&			&	19.1	&	56.00	&	7.5		&	5.4	&	0.51	&	-63.3	&	19.7	&	44.00	&	8.9		&	0.3	&	0.80	&	-82.5	\\
J031927.22+000014.5 &	1		&	18.8	&	4.5		&	   		&			&	18.7	&	100.00	&	7.4		&	3.8	&	0.64	&	33.8	&	   		&			&			&		&			&		\\
J032029.78+003153.5 &	2C		&	17.0	&	12.8	&	21.6	&	2.99	&	19.3	&	23.71	&	3.4		&	4.0	&	0.85	&	63.9	&	17.8	&	73.3	&	28.9	&	1.0	&	0.51	&	72.7	\\
J034215.08+001010.6 &	1		&	19.2	&	1.1		&	   		&			&	19.1	&	100.00	&	0.7		&	6.0	&	0.82	&	-14.7	&	   		&			&			&		&			&		\\
J040152.38-053228.7 &	2B		&	19.0	&	2.7		&	21.1	&	12.17	&	18.9	&	87.83	&	2.9		&	1.0	&	0.87	&	-40.1&	   		&			&			&		&			&				\\
J074811.44+395238.0 &	2B		&	18.7	&	4.6		&	21.3	&	9.45	&	18.6	&	90.55	&	9.4		&	4.0	&	0.89	&	-70.5	&	   		&			&			&		&			&		\\
J081125.81+073235.3 &	1		&	19.2	&	4.0		&	   		&			&	19.2	&	100.00	&	5.9		&	4.2	&	0.73	&	-0.9	&	   		&			&			&		&			&		\\
J082449.27+370355.7 &	2B		&	19.0	&	1.3		&	21.1	&	14.20	&	19.1	&	85.80	&	2.0		&	5.7	&	0.72	&	62.5	&	   		&			&			&		&			&		\\
J082527.50+202543.4 &	2B		&	19.7	&	1.1		&	20.4	&	49.48	&	20.4	&	50.52	&	1.5		&	1.0	&	0.95	&	-64.5	&	   		&			&			&		&			&			\\
J083028.14+202015.7 &	2C		&	18.5	&	2.5		&	20.8	&	11.85	&	19.2	&	48.83	&	1.8		&	4.0	&	0.33	&	-38.6	&	19.4	&	39.32	&	3.1		&	1.0	&	0.95	&	11.2	\\
J084041.08+383819.8 &	2C		&	17.8	&	7.1		&	20.8	&	6.46	&	19.7	&	18.47	&	4.2		&	0.4	&	0.33	&	5.2		&	18.1	&	75.08	&	9.5		&	1.0	&	0.81	&	-28.3	\\
J084309.86+294404.7 &	2B		&	18.7	&	3.1		&	20.3	&	22.02	&	19.0	&	77.98	&	6.7		&	2.6	&	0.87	&	-49.0		&	   		&			&			&		&			&		\\
J084856.58+013647.8 &	1		&	17.1	&	4.9		&	   		&			&	17.9	&	100.00	&	18.7	&	5.0	&	0.87	&	-39.6	&	   		&			&			&		&			&		\\
J084943.82+015058.2 &	2C		&	19.5	&	2.2		&	21.2	&	21.57	&	20.2	&	44.41	&	3.8		&	4.0	&	0.41	&	66.4	&	20.7	&	34.03	&	3.1		&	1.0	&	0.47	&	50.8	\\
J090414.10-002144.9 &	2B		&	17.8	&	2.9		&	21.3	&	9.90	&	18.3	&	90.10	&	9.6		&	6.5	&	0.52	&	-67.3	&	   		&			&			&		&			&		\\
J092318.06+010144.8 &	1		&	18.8	&	3.1		&	   		&			&	18.8	&	100.00	&	4.1		&	4.0	&	0.68	&	-46.5	&	   		&			&			&		&			&		\\
J094209.00+570019.7 &	2D		&	18.7	&	1.6		&	19.9	&	31.42	&	19.4	&	53.12	&	4.3		&	5.2	&	0.91	&	-54.4	&	20.7	&	15.46	&	4.3		&	0.2	&	0.75	&	-86.3	\\
J094350.92+610255.9 &	2B		&	19.6	&	1.0		&	20.2	&	49.77	&	20.3	&	50.23	&	1.4		&	1.0	&	0.85	&	-48.1	&	   		&			&			&		&			&			\\
J095629.06+573508.9 &	2B		&	19.0	&	3.2		&	21.3	&	13.44	&	19.1	&	86.56	&	4.5		&	2.2	&	0.58	&	-80.5	&	   		&			&			&		&			&		\\
J100329.86+511630.7 &	1		&	18.4	&	4.8		&	   		&			&	18.4	&	100.00	&	6.3		&	2.5	&	0.48	&	75.8	&	   		&			&			&		&			&		\\
J103639.39+640924.7 &	1		&	17.8	&	7.3		&	   		&			&	17.8	&	100.00	&	14.0	&	4.0	&	0.65	&	19.6	&	   		&			&			&		&			&		\\
J112907.09+575605.4 &	2D		&	19.1	&	2.4		&	21.8	&	9.38	&	19.2	&	73.34	&	5.1		&	4.0	&	0.70	&	-5.2	&	21.0	&	17.28	&	1.8		&	0.2	&	0.07	&	-7.6	\\
J113710.78+573158.7 &	1		&	18.6	&	3.3		&	   		&			&	18.6	&	100.00	&	7.5		&	6.5	&	0.53	&	-48.3	&	   		&			&			&		&			&		\\
J140740.06+021748.3 &	2D		&	19.2	&	1.5		&	   		&			&	19.3	&	86.28	&	1.2		&	5.6	&	0.46	&	-45.3	&	21.2	&	13.72	&	5.2		&	0.5	&	0.15	&	-35.8	\\
J150117.96+545518.3 &	2B		&	17.2	&	6.1		&	19.5	&	11.39	&	17.2	&	88.61	&	8.5		&	2.3	&	0.88	&	62.3	&	   		&			&			&		&			&		\\
J154133.19+521200.1 &	2A		&	18.0	&	6.7		&	   		&			&	19.5	&	25.59	&	2.0		&	4.6	&	0.65	&	-29.7	&	18.1	&	74.41	&	9.8		&	1.0	&	0.79	&	-19.7	\\
J154613.27-000513.5 &	2B		&	19.0	&	3.8		&	21.3	&	12.71	&	19.1	&	87.29	&	5.9		&	2.4	&	0.84	&	-51.2	&	   		&			&			&		&			&		\\
J172419.89+551058.8 &	2B		&	19.3	&	1.4		&	20.3	&	43.86	&	19.9	&	56.14	&	5.0		&	1.7	&	0.63	&	63.9	&	   		&			&			&		&			&		\\
J172603.09+602115.7 &	2B		&	19.8	&	1.2		&	20.6	&	47.06	&	20.3	&	52.94	&	2.6		&	1.4	&	0.52	&	45.4	&	   		&			&			&		&			&		\\
J173938.64+544208.6 &	2C		&	19.2	&	2.6		&	20.9	&	20.51	&	20.3	&	35.77	&	1.9		&	0.8	&	0.54	&	37.9	&	20.1	&	43.72	&	8.5		&	1.0	&	0.51	&	33.4	\\
J214415.61+125503.0 &	2C		&	18.5	&	1.2		&	19.0	&	50.67	&	19.8	&	11.05	&	16.6	&	5.4	&	0.32	&	-32.9	&	19.2	&	38.29	&	5.1		&	1.0	&	0.73	&	-72.2	\\
J223959.04+005138.3 &	2C		&	18.9	&	3.6		&	20.5	&	22.12	&	20.7	&	19.38	&	2.1		&	0.3	&	0.74	&	-1.8	&	19.2	&	58.50	&	9.3		&	1.0	&	0.50	&	61.8	\\
J231755.35+145349.4 &	2B		&	18.3	&	6.2		&	22.1	&	3.10	&	18.3	&	96.90	&	9.7		&	2.0	&	0.60	&	79.1	&	   		&			&			&		&			&		\\
J231845.12-002951.4 &	2D		&	18.4	&	4.3		&	20.1	&	9.85	&	18.1	&	77.73	&	24.7	&	4.8	&	0.78	&	48.6	&	20.6	&	12.41	&	7.2		&	0.3	&	0.20	&	23.4	\\ \hline
 \end{tabular}
\caption{List of the 42/57 objects that could be fitted with {\sc GALFIT} and results of the best fits. {\b Col(2): Method to obtain the best fit. 1: Single S\'ersic. 2A:S\'ersic+Disk, 2B: PS+S\'ersic, 2C: PS+Disk+S\'ersic, 2D:Other.} Col(3,4):  Total magnitude and effective radius $R_{\rm eff }$ in kpc of the {\sc GALFIT}  galaxy model. Col(5,6):   Magnitude and  relative contribution (light fraction)  to the total galaxy luminosity of the point source.   Col(7 to 12): Magnitude, light fraction and structural parameters of 1st component identified with {\sc GALFIT}. Col(13 to 18): same for 2nd component.   S\'ersic index $n$, axis ratio $b/a$, position angle (east of north) $PA$.}
\label{Table:param}
	\end{table}
\end{landscape}

\begin{landscape}
\begin{table}
\begin{tabular}{|l|c|c|c|c|c|c|c|}
\hline \hline
			&			 & \multicolumn{3}{c|}{This work} 	&\multicolumn{2}{c|}{B/T Classification}&		Visual vs. Parametric	\\  \cmidrule(l){3 - 5} \cmidrule(l){6 - 7} 
\multicolumn{1}{c}{Object SDSS}	&	$lO3$	&  Vis-I &	Vis-II	&	Vis-III	&	Method	&	Details	&		\\ 
\multicolumn{1}{c}{[1] }	&	[2]	&	[3]		&	[4]	& [5]				&	[6]	&	[7]	    			&	[8]		    	\\  \hline
J002531.46-104022.2	&	8.73	&   HD		&	1   	&  2N, T, K?		&	N/A	&	HD			&		\\
J005515.82-004648.6	&	8.15	& 	El		&	0    	&				&	2B	&	Disk-Dominated	&	El vs. Disk-Dominated	\\
J011429.61+000036.7	&	8.66	& 	El		&	0    	&  	PC			&	2B	&	Bulge-Dominated	&		\\
J011522.19+001518.5	&	8.14	& 	El		&	0*	&  A,  tt, PC		&	1 	&	Bulge-Dominated	&		\\
J014237.49+144117.9	&	8.76	& 	El		&	0*	&PC, S,D, tt, K?	&      2B	&	Bulge-Dominated	&		\\
J015911.66+143922.5	&	8.56	& 	El		&	0    	&				&	2B	&	Disk-Dominated	&	El vs. Disk-Dominated		\\
J020234.56-093921.9	&	8.39	& 	Sp		&	1    	&	IC,B, K?		&	2A	&	Disk-Dominated		&	\\
J021059.66-011145.5	&	8.10	& 	Sp		&   0*	&	S			&	2B	&	Bulge-Dominated	&	Sp vs. Bulge-Dominated \\
J021758.18-001302.7	&	8.55	&   HD		&	2    	&	T,D			&	N/A	&	HD				&		\\
J021834.42-004610.3	&	8.85	& 	HD		&	1    	&	T,C,IC = 2N	&	N/A	&	HD				&		\\
J022701.23+010712.3	&	8.90	& 	HD		&	2    	&2 N, T, D, S , A	&	N/A	&	HD				&		\\
J023411.77-074538.4	&	8.77	& 	El		&	0    	&				&	2D	&	Bulge-Dominated	&		\\
J031946.03-001629.1	&	8.24	& 	Sp or Disk	&	0    	&				&	2D	&	Bulge-Dominated	&	 Sp or Disk vs. Bulge-Dominated\\
J031927.22+000014.5	&	8.06	& 	El		&	0*	&	F, tt?			&	1 	&	Bulge-Dominated	&		\\
J032029.78+003153.5	&	8.52	& 	El		&	0*	&	D, PC		&	2C 	&	Disk-Dominated	&	El vs. Disk-Dominated		\\
J032533.33-003216.5	&	9.06	& 	El		&	2    	&	T			&	N/A	&	HD			&	El vs. HD\\
J033310.10+000849.1	&	8.13	& 	HD		&	1    	&  2N/D?, T, K?, IC	&	N/A	&	HD			&		\\
J034215.08+001010.6	&	9.08	& 	El		&	0    	&				&	1 	&	Bulge-Dominated	&		\\
J040152.38-053228.7	&	8.96	& 	El		&	0*	&	A, PC?		&	2B	&	Disk-Dominated	&	El vs. Disk-Dominated		\\
J074811.44+395238.0	&	8.19	& 	El		&	2    	&	T, K?			&	2B	&	Bulge-Dominated	&		\\
J081125.81+073235.3	&	8.88	& 	El		&	0*	&  A, IC?, I		&	1 	&	Bulge-Dominated	&		\\
J081330.42+320506.0	&	8.83	& 	El?		&	2    	&	T,tt, D, I, A 	&	N/A	&	HD			&	El? vs. HD	\\
J082449.27+370355.7	&	8.28	& 	El		&	0    	&				&	2B	&	Bulge-Dominated	&		\\
J082527.50+202543.4	&	8.88	& 	El		&	0    	&				&	2B	&	Disk-Dominated	&	El vs. Disk-Dominated		\\
J083028.14+202015.7	&	8.91	& 	El		&	0    	&				&	2C 	&	Bulge-Dominated	&		\\
J084041.08+383819.8	&	8.47	& 	Sp		&	2    	&	T, K?			&	2C	&	Disk-Dominated	&		\\
J084309.86+294404.7	&	9.34	& 	El		&	0    	&				&	2B	&	Bulge-Dominated	&		\\
J084856.58+013647.8	&	8.46	& 	El 		&	0*	&	S, A, PC		&	1 	&	Bulge-Dominated	&		\\  \hline
 \end{tabular}
 \caption{Results of the visual and parametric classification (see Section \ref{Class:methods} for details): Col(3) indicates the galaxy morphological type based on Method Vis-I:  El (elliptical), Sp (spiral), Disk (disk galaxy with no obvious spiral arms), HD (highly-disturbed: when 2 or more components are interacting and make the classification in  the previous groups difficult).  Col(4):  type of merger based on  Method Vis-II following  \citet{Rodriguez2011} and \citet{Veilleux2002}: 0= Isolated undisturbed galaxy, 0*= Isolated disturbed galaxy, 1= two  nuclei with projected separation  $>$ 1.5 kpc;  2 = two nuclei  with projected separation $\leq 1.5$ or single nucleus with signatures of a post-coalescence phase. Col(5) specify  the nature of the peculiar features based on Method Vis-III  following \citet{CRamos2011}. T: Tidal tail; F: Fan;  B: Bridge; S: Shell; D: Dust feature; 2N: Double Nucleus;  A: Amorphous Halo; I: Irregular feature. IC: interacting companion. Other features we include in this work are 2N*: two nuclei with relative distance  greater than 1.5 kpc, PC: projected companion, this is,  there is an object close to the target in projection with no clear physical relation, K: Knot, tt: Streams. A question mark ''?'' indicates uncertain classification or identification.
 Col(6): classification based on the parametric method  (Table \ref{Table:param}). N/A refers to objects that could not be fitted with {\sc GALFIT}. Col(7): type of profile according to the dominant structural component:	HD in this column are highly-disturbed objects that could not be fitted with {\sc GALFIT}.  The vast majority are highly-disturbed systems. Point source: system  dominated by a spatially unresolved source; Disk-dominated: system dominated by a disk or disk-like component; 
Bulge-dominated: system dominated by a spheroidal component.
 Col(8): Comparison between the visual   and  parametric classifications.  }
\label{Table:class}
\end{table}
\end{landscape}

\addtocounter{table}{-1}

\begin{landscape}
\begin{table}
\begin{tabular}{|l|c|c|c|c|c|c|c|}
\hline \hline
			&			 & \multicolumn{3}{c|}{This work} 	&\multicolumn{2}{c|}{B/T Classification}&	Visual vs. Parametric	\\  \cmidrule(l){3 - 5} \cmidrule(l){6 - 7} 
\multicolumn{1}{c}{Object SDSS}	&	$lO3$	&  Vis-I &	Vis-II	&	Vis-III	&	Method	&	Details	&		\\ 
\multicolumn{1}{c}{[1] }	&	[2]	&	[3]		&	[4]	& [5]				&	[6]	&	[7]	    			&	[8]		    	\\  \hline
J084943.82+015058.2	&	8.06	&	El			&	0    	&				&	2C  	&	Bulge-Dominated		&	 	\\
J090307.84+021152.2	&	8.42	&      HD			&	 1/2	& 2N*, T, K, D		&	N/A	&	HD			&		\\
J090414.10-002144.9	&	8.93	&	El			&	2    	&	T, F			&	2B 	&	Bulge-Dominated	&		\\
J090801.32+434722.6	&	8.31	&	El			&	2    	&	T, I, K		&	N/A	&	HD				&	El vs. HD	\\
J092318.06+010144.8	&	8.94	&	Sp			&   0*	&	T? 			&	1 	&	Bulge-Dominated	&	Sp vs B \\
J092356.44+012002.1	&	8.59	&	El			&	 1/2	&	T			&	N/A	&	HD				&	El vs. HD	\\
J094209.00+570019.7	&	8.31	&	El			&	1    	& T, D, F, tt, PC		& 	2D 	&	Bulge-Dominated	&		\\
J094350.92+610255.9	&	8.46	&	El			&	0    	&				&	2B 	&	Point Source		&	El vs. Point Source	\\
J095629.06+573508.9	&	8.38	&	El			&	0    	&				&	2B 	&	Bulge-Dominated	&		\\
J100329.86+511630.7	&	8.11	&	El			&	2    	&	T, A, F		&	1  	&	Bulge-Dominated	&		\\
J103639.39+640924.7	&	8.42	&	El			&	2    	&	T, PC		&	1  	&	Bulge-Dominated	&		\\
J112907.09+575605.4	&	9.38	&	El			&	0*	&	I			&	2D 	&	Bulge-Dominated	&		\\
J113710.78+573158.7	&	9.61	&	El 			&	0*	&	T, tt			&	1  	&	Bulge-Dominated	&		\\
J133735.01-012815.7	&	8.72	&	El			&	2    	&	T?, 2N, I		&	N/A	&	HD				&	El vs. HD \\
J140740.06+021748.3	&	8.90	&	Sp or Disk		&	0    	&				&	2D 	&	Bulge-Dominated	&	Sp or Disk vs. Bulge-Dominated	\\
J143027.66-0056149	&	8.44	&	HD			&	1    	& 2N, T, IC?, K, A	&	N/A	&	HD				&		\\
J144711.29+021136.2	&	8.45	&      HD			&	 1/2	& 2N*, T, B		&	N/A	&	HD				&		\\
J150117.96+545518.3	&	9.06	&	El?			&	1    	& IC, S,D, T, B, K?	&      2B 	&	Bulge-Dominated	&		\\
J154133.19+521200.1	&	8.25	&	Sp			&	2    	&	S, K, T, IC?	&	2A 	&	Disk-Dominated	&		\\
J154337.81-004420.0	&	8.40	&	HD			&	1    	&	2N, T,F,I, A	&	N/A	&	HD				&		\\
J154613.27-000513.5	&	8.18	&	El			&	0    	&				&	2B 	&	Bulge-Dominated	&		\\
J172419.89+551058.8	&	8.00	&	El			&	0    	&			&	2B 	&	Disk-Dominated	&	El vs. D	\\
J172603.09+602115.7	&	8.57	&	El			&	0    	&				&	2B 	&	Disk-Dominated	&	El vs. Disk-Dominated	\\
J173938.64+544208.6	&	8.42	&	Sp or Disk		&	0*	&	tt?			&	2C 	&	Disk-Dominated	&		\\
J214415.61+125503.0	&	8.14	&	El			&	2    	&	I,T			&	2C	&	Point Source	&	El vs. Point Source	\\
J215731.40+003757.1	&	8.39	&	HD			&	1    	&	D,IC			&	N/A	&	HD				&		\\
J223959.04+005138.3	&	8.15	&	Sp or Disk		&	0*	&	A?			&	2C 	&	Disk-Dominated	&		\\
J231755.35+145349.4	&	8.10	&	Sp			&	0    	&	C			&	2B 	&	Bulge-Dominated	&	Sp vs. Bulge-Dominated	\\
J231845.12-002951.4	&	8.00	&	Sp			&	0    	&				&	2D 	&	Bulge-Dominated	&	Sp vs. Bulge-Dominated	\\  \hline \end{tabular}
 \caption{\bf (Cont.)}
 \label{Table:5:b}
 \end{table}
 \end{landscape}
\newpage

 Some of these  works have suggested that the type of galaxy host depends on AGN power, so   that the most luminous AGN (quasars) tend to be  hosted by massive ellipticals (\citealt{Dunlop2003}, \citealt{Floyd2004}). This is however controversial and other works claim a higher incidence of galaxies with disks  in luminous quasars (e.g. \citealt{Cales2011}, \citealt{Falomo2014}). 

The  contribution of this work to the topic of the host galaxies associated with quasars is based on: 1) due to their recent discovery, just a few studies exist (two, to the best of our knowledge:  \citealt{Greene2009}, \citealt{Wylezalek2016})  regarding the structural properties of $z<$1 QSO2 host galaxies . They have been focussed on  samples of up to $\sim$20  objects. We expand these works with 41 more QSO2 and complement it with 16 HLSy2; 2) we take into account factors that sometimes have not been considered, such as the high incidence of highly-disturbed systems,  the contribution of a point source and/or the presence  of several structural components in galaxies;   3) we perform a thorough comparative study with related parametric works  of the host galaxies of luminous AGN at similar $z$ (see Sect. \ref{Analogous:Studies}).

Our analysis has shown a wide diversity of galaxy hosts among both HLSy2 and  QSO2 (Tables \ref{compa-merger} and \ref{compa}). An interesting result is that a  high fraction ($\sim$55\%)    of QSO2 are {\it not} hosted by bulge-dominated galaxies. Although  these are the most numerous group  (44$\%^{+5}_{-10}$), more than half QSO2 are hosted by other galaxy types, mostly   highly-disturbed systems due to galaxy interactions  (34$\%^{+6}_{-9}$) and disk-dominated systems (20$\%^{+6}_{-6}$).  The main difference between QSO2 and HLSy2 is the lower incidence of morphologically-disturbed systems among HLSy2 (6$\%^{+10}_{-3}$). This is consistent with a scenario in which galaxy interactions are the dominant mechanism triggering the activity at the highest AGN power (e.g. \citealt{Hopkins2008,Bessiere2012, CRamos2011}). 

Although disks are identified in 44$\%^{+13}_{-11}$ of HLSy2 and 24$\%^{+8}_{-6}$ QSO2, disk-dominated systems represent a minority (25$\%^{+8}_{-11}$  HLSy2 and 20$\%^{+6}_{-6}$ QSO2).  Excluding highly-disturbed systems, if we consider the coarse division QSO2 vs. HLSy2 at $lO3=8.3$, both groups show a similar distribution among galaxy types (Tables \ref{compa-merger} and \ref{compa}). However, a more careful analysis reveals that the galaxy properties do change with AGN power:  the relative contribution of the spheroidal component to the total galaxy light (B/T) increases with $lO3$  (Sec. \ref{bt}). Excluding the complex merger/interaction systems, B/T$\ga$0.7 for most galaxies with   $lO3\ga$8.6. As other authors have argued, this is naturally expected if more powerful AGN are powered by more massive black holes which in turn are hosted by more massive bulges or spheroids (\citealt{Magorrian1998}, \citealt{Dunlop2003}). Constraining the galaxy masses of the sample studied here would very valuable to investigate this scenario.

\begin{center}
\begin{table*}
\footnotesize
\begin{tabular}{|l|c|c|c|c|c|}
\hline 
					&		Disk-dominated	   	   &	Bulge-dominated	   &		B+D*		  &	 Point Source 	   &		Disturbed		\\	 \hline
This work		(HLSy2)		& $	25\%^{+8}_{-11}	$ & $	63\%^{+1}_{-21}	$ & $0\%^{+9}$ & $ 6\%^{+10}_{-3}$ & $   6\%^{+10}_{-3}   $ \\	 \hline
This work		(QSO2)		& $	20\%^{+6}_{-6}		$ & $	44\%^{+5}_{-10}	$ & $0\%^{+4}$ & $ 2\%^{+5}_{-0.9}$ & $   34\%^{+6}_{-9}   $ \\	 \hline \hline
Dunlop et al. (2003)	(RL-QSO1)	& 				   & $100\%_{-16}   		$ & 				  &  				  &					\\	\hline
Dunlop et al. (2003)	(RQ-QSO1)	& $15\%^{+15}_{-6}	$ & $85\%^{+6}_{-15}	$ & 				  &  				  &					\\	\hline
Floyd et al. (2004)	(QSO1)	& $	18\%^{+11}_{-7}	$ & $	76\%^{+8}_{-13}	$ & $6\%^{+11}_{-3}	$ &  				   &		 	 		\\	 \hline
Falomo et al. (2014)	 (QSO1)	& $	42\%^{+3}_{-3}		$ & $	37\%^{+3}_{-3}		$ & $21\%^{+3}_{-3}	$ &			  	   & 					\\	 \hline
Cales et al. (2011)	(QSO1) 	& $	21\%^{+8}_{-7}		$ & $62\%^{+4}_{-14} 	$ & $	3\%^{+7}_{-1.2}$ & 				   & $	  14\%^{+8}_{-5} $     \\	 \hline
Greene et al. (2009)	 (QSO2)	& $	20\%^{+12}_{-8}	$ & $	53\%^{+9}_{-16}	$ &  				   &  				   & $27\%^{+12}_{-10}$	 \\	 \hline
Wylezalek et al. (2016) (QSO2)& $   10\%^{+11}_{-4}	$ & $90\%^{+4}_{-11}	$ & 				   &  				   &					 \\	 \hline
Inskip et al. (2010)	(NLRG)	& $ 	6\%^{+11}_{-3}		$ & $88\%^{+1}_{-17}	$ & $ 6\%^{+11}_{-3}	$ &  				   &  					 \\	 \hline
\end{tabular}
\caption{Comparison with other works. $^*$B/D  include intermediate classification in works that use only one S\'ersic profile and $n$ has an intermediate value between disk-dominated and bulge-dominated objects asummed by the authors.}
\label{compa-merger}
\end{table*}
\end{center}

\begin{center}
\begin{table*}
\footnotesize
\begin{tabular}{|l|c|c|c|}
\hline 
						&		Disk-dominated		   &		Bulge-dominated	   &		B+D*		 \\	 \hline	 	
This work		(HLSy2)		& $	29\%^{+12}_{-10}	$ & $	71\%^{+5}_{-19}	$ & $ 0\%^{+11}	$ \\	 \hline	
This work		(QSO2)		& $	31\%^{+9}_{-9}		$ & $	69\%^{+6}_{-13}	$ & $ 0\%^{+7}$ \\	 \hline	\hline
Dunlop et al. (2003)	(RL-QSO1)	& 				   & $100\%_{-16}   		$ & 				  	  \\	\hline
Dunlop et al. (2003)	(RQ-QSO1)	& $15\%^{+15}_{-6}	$ & $85\%^{+6}_{-15}	$ & 				          \\	\hline
Floyd et al. (2004)	(QSO1)	& $	18\%^{+11}_{-7}	$ & $	76\%^{+8}_{-13}	$ & $	  6\%^{+11}_{-3}	$ \\	 \hline	
Falomo et al. (2014)	 (QSO1)	& $	42\%^{+3}_{-3}		$ & $	37\%^{+3}_{-3}		$ & $	  21\%^{+3}_{-3}	$ \\	 \hline
Cales et al. (2011)	(QSO1)	& $	24\%^{+9}_{-7}		$ & $	72\%^{+5}_{-13}	$ & $	  4\%^{+8}_{-1.4}	$ \\	 \hline	
Greene et al. (2009)	 (QSO2)	& $	27\%^{+19}_{-9}	$ & $	73\%^{+9}_{-17}	$ & $					$ \\	 \hline
Wylezalek et al. (2016) (QSO2)	& $   10\%^{+11}_{-4}	$ & $90\%^{+4}_{-11}	$ & $					$ \\	 \hline
Inskip et al. (2010)	 (NLRG)	& $	6\%^{+11}_{-3}		$ & $	88\%^{+1}_{-17}	$ & $  6\%^{+11}_{-3}	$ \\	 \hline
\end{tabular}
\caption{As Table  6 but excluding highly-disturbed systems.}
\label{compa}
\end{table*}
\end{center}

\subsection{Comparison with other works}
\label{Analogous:Studies}

We put the results of our parametric analysis in the context of other relevant works. We focus our comparison on the QSO2 sub-sample and related studies of AGN  with quasar like luminosities (QSO1, QSO2 and NLRG). While our HLSy2 are at the high end of Sy2  luminosities, related studies on Seyferts cover a much wider range usually extending to significantly lower AGN power, so that the comparison is not trivial \citep{Kauffmann2003}. 

We  also focus on studies based on samples at $z\la$0.5.  The comparison with high $z$ studies is complicated by  the  limited physical spatial information and/or the shallowness of the data and/or the different rest-frame spectral range (e.g. rest-frame UV at $z>2$  versus rest-frame optical at low $z$). Detailed information on all  referenced works  can be found in Appendix \ref{otherworks}.

Firm conclusions regarding the origin of some discrepancies and similarities between works are not possible due to the numerous potential influencing factors on the galaxy classification: poor statistics, data properties (depth, spectral range, spatial resolution), sample selection (range of AGN luminosities, radio-loudness, obscured versus unobscured), fitting method (e.g. one versus several structural components; classification criteria based on $n$ values).  For the sake of clarity,  
 we mention for each work the available information that can help the reader identify the possible influencing factors (Appendix \ref{otherworks}).  The results of all works are summarized in Tables   \ref{compa-merger} and  \ref{compa}.
The confidence intervals have been calculated as in Sect. \ref{Visual:classification:1}. 

In spite of the above limitations,  some interesting results appear, which can be summarized as follows.

\begin{itemize}
 \item A clear difference between works is the incidence of highly-disturbed merger/interaction systems, which are absent in several works while they account for   $\sim$34\% of our QSO2 sample. The reason for the discrepancy lies, at least in part, in that  the classification and/or fitting methods are often not sensitive to the distinction of such systems. 

\item  In general, all works are consistent regarding the fraction of disk-dominated galaxies ($\sim$10-20\%) and B+D systems ($\la$few \%) in radio-quiet quasars.  This fraction is tentatively  lower in radio-loud QSO1 and NLRG ($\la$6\%, \citealt{Dunlop2003,Inskip2010}).  
 This is consistent with the fact that  powerful radio-loud objects  tend to be hosted by  massive elliptical galaxies \citep{Matthews1964,Best2005}.

\citet{Falomo2014}  is an exception.
 They find a significantly higher fraction of disk-dominated systems (42\%$^{+3}_{-3}$) and B+D (21\%$^{+3}_{-3}$). They identify disks in a high fraction of quasars $\sim$63\% (e.g. $\sim$24\% in our sample).  A real  difference in the type of galaxy hosts  cannot be discarded, but it must be kept it mind that the  fitting and classification method applied are, as the authors  warn,  too simplistic  and can only yield a preliminary indication of the morphology (Appendix \ref{otherworks}). 

\item All works are consistent in that the most numerous group of host galaxies are always bulge-dominated  (\citealt{Falomo2014} is again an exception). On the other hand, the percentage  varies significantly.  In general,  studies that do not separate complex merger/interaction systems   (\citealt{Dunlop2003,Floyd2004,Inskip2010,Wylezalek2016}) result on a higher fraction of bulge-dominated galaxies ($\sim$76-100\%)  compared with other works where disturbed systems are identified (44\%-62\%, our work, \citealt{Greene2009,Cales2011}).  Thus, the classification criteria may play a role on the observed differences. 

\end{itemize}

Moreover, intrinsic differences between samples probably also play a role. For instance, the radio-loud (\citealt{Dunlop2003,Inskip2010}) and most luminous  samples  \citep{Wylezalek2016} show tentative evidence for the highest fraction of bulge-dominated systems. This is naturally expected. On one hand,  radio loudness is favoured in  massive bulge dominated  galaxies, as mentioned above. On the other hand,  as we have seen, the relative contribution of the spheroidal component to the total galaxy light increases with $lO3$, proxy of AGN power (Sect. \ref{bt}).

\section{Summary and conclusions}
\label{Conclusions}

We have studied the morphological and parametric properties of  the host galaxies of 57 optically-selected luminous type 2 AGN at $0.3 \la z \la 0.4$ from the SDSS.  The sample consists of  41  QSO2 with 8.31$\le lO3\le$9.61 and 16 high-luminosity Seyfert 2  (HLSy2) with 8.06$\le lO3\le$8.28. 
 Our study is based on  HLA archive ACS/WFC and WFPC2  HST images. Both samples contain $\sim$44\% of all SDSS optically selected QSO2 and HLSy2 within the same $z$ and $L_{\rm [OIII]}$ ranges. Although uncertainties remain regarding the exact selection criteria, we consider them an adequate representation of the original total SDSS samples.

Due to the recent discovery of QSO2 in large numbers,  the structural properties of their host galaxies are poorly known. To our knowledge, only two related  studies have been published for $z<$1 QSO2, with 35 QSO2 hosts parametrized so far ( \citealt{Greene2009}, \citealt{Wylezalek2016}).  Our work expands this investigagion with 41 more QSO2 and complements it with 16 HLSy2.

We have  classified the galaxies both visually and, most importantly, parametrically using the code {\sc GALFIT}.  The parametric analysis is essential to isolate and parametrize the individual structural galactic components  and, ultimately, classify the galaxies in terms of the dominant  component. 

The main results and conclusions of our study are:

 \begin{itemize}

\item  There is a wide diversity of galaxy hosts among both HLSy2 and  QSO2.   Less than half  (44$\%^{+5}_{-10}$) of QSO2 are  hosted
by bulge-dominated galaxies (in our terminology this includes spheroidal galaxies and galaxies with B/D$>$1.2). More than half  are hosted by other galaxy types, mostly   highly-disturbed systems due to galaxy mergers/interactions  (34$\%^{+6}_{-9}$) and disk-dominated systems (20$\%^{+6}_{-6}$, these are disk like galaxies and galaxies with B/D$<$0.8). A minority of galaxies  are dominated by a point  source (2$\%^{+5}_{-0.9}$)

\item Among HLSy2,  63$\%^{+1}_{-21}$ are bulge-dominated, 25$\%^{+8}_{-11}$  are disk-dominated and 6$\%^{+10}_{-3}$  are highly-disturbed systems. 6$\%^{+10}_{-3}$ are dominated by a point source.

\item A significant difference between QSO2 and HLSy2 is the higher incidence of morphologically disturbed systems among QSO2 (34$\%^{+6}_{-9}$ versus 6$\%^{+10}_{-3}$). This is consistent with a scenario in which galaxy interactions are the dominant mechanism triggering the activity at the highest AGN power.

\item Disks are identified in a significantly higher fraction of HLSy2  (44$\%^{+13}_{-11}$) than QSO2 (24$\%^{+8}_{-6}$) but this is a result of the lower fraction of disturbed systems among Seyferts. When these are not considered, the fractions become consistent within the errors  (47$\%^{+13}_{-12}$  HLSy2 and  37$\%^{+10}_{-8}$ QSO2). 

\item The coarse and somewhat arbitrary division between HLSy2 and QSO2 at $lO3=8.3$  is not adequate to unveil trends of galaxy host with $lO3$ (used here as proxy of AGN power). A more detailed analysis using a finer sampling of  $lO3$, reveals a clear dependence of the galaxy properties with  AGN power. The relative contribution of the spheroidal component to the total galaxy light (B/T) increases with $L_ {\rm [OIII]}$.  B/T$\ga$0.7 for most QSO2 with   $lO3\ga$8.6, while at lower luminosities the galaxies span the full range  B/T$\sim$0.0-1.0.   While bulge-dominated systems spread across the total range of $L_ {\rm [OIII]}$  of the sample, most disk-dominated galaxies concentrate at $lO3$$\la$8.6.

As other authors have argued, this is naturally expected if more powerful AGN are powered by more massive black holes which in turn are hosted by more massive bulges or spheroids. Constraining the galaxy masses of the sample studied here would be very valuable to investigate this scenario.

 \item A point source component is isolated in a high fraction of objects (10/16 or $63\%^{+11}_{-13}$ HLSy2 and 19/41 or $\sim46\%^{+8}_{-8}$ QSO2), although it rarely dominates the total flux. The relative contribution  to the total galaxy light   is in the range $\sim$3-51\% (median value 14.2\%). 
In spite of being obscured AGN, even when the light fraction is small ($\sim$few\%), this point source cannot be ignored in the fits, since the structural parameters of the hosts can be severely affected.

\item We have compared our results with other works dedicated to the parametric classification of the host galaxies of luminous AGN in the quasar regime (QSO, QSO2 and NLRG).  All works are in general consistent regarding the fraction of disk-dominated galaxies in radio-quiet QSO1 and QSO2 ($\sim$10-20\%). This fraction appears to be lower in radio-loud systems. This is consistent with the fact that powerful radio-loud AGN tend to be hosted by massive ellipticals.  All works are in general also consistent in that bulge-dominated systems are the most numerous, although the percentages vary significantly among works. This is due to possible difference between samples (e.g. radio-loud versus radio-quiet, range of AGN power), differences in the parametric classification method and the misclassification of highly-disturbed systems.

\item  The spheroidal component of the host galaxies of the sample studied here   follows the Kormendy relation defined by   early-type galaxies at $0.01\le z\le$0.3.
The  slope is also consistent with the relation found for the bulge-dominated host galaxies
    of QSO1 at similar $z$. The average galaxy sizes (correcting for PSF contamination)   are 
$\langle r_{\rm e} \rangle$=3.9$\pm$0.6 kpc (median 3.3$\pm$1.6 kpc) for  QSO2 and  5.0$\pm$1.5 kpc (median 4.8$\pm$0.9 kpc)  for HLSy2. The QSO2 sizes are consistent with  those expected for non active galaxies at $z\sim$0 of  stellar masses in the range expected for our sample. They are, on the other hand, smaller than $\langle r_{\rm e} \rangle \sim$7-10 kpc quoted in the literature for QSO1 and NLRG. The latter samples may be biased towards more massive galaxies.

 \end{itemize}

\section*{Acknowledgments}

Thanks to an anonomous referee for useful comments on and suggestions on the paper.  JUM and MVM acknowledge  support  from the  Spanish former Ministerio de Econom\'\i a y Competitividad through the grants  AYA2012-32295 and AYA2015-64346-C2-2-P. FB acknowledges the support by FCT via the postdoctoral fellowship SFRH/BPD/103958/2014. This work is supported by Fundacao para a Ciencia e a Tecnologia (FCT) through national funds (UID/FIS/04434/2013) and by FEDER through COMPETE2020 (POCI-01-0145-FEDER-007672). FB also acknowledges support from grant AYA2016-77237-C3-1-P from the Spanish Ministry of Economy and Competitiveness (MINECO). JPL acknowledges support from the Spanish Ministerio de Econom\'\i a y Competitividad through the grant AYA2017-85170-R. BRP acknowledges financial support from the Spanish Ministry of Economy and Competitive- ness through grant ESP2015-68964. F.J.C. acknowledges
financial support through grant AYA2015-64346-C2-1-P (MINECO/FEDER).

We thank Enrica Bellocchi  for useful scientific discussions and Boris H\"aussler and Chien Peng for valuable advice on the use of {\sc GALFIT}.

Based on observations made with the NASA/ESA Hubble Space Telescope, and obtained from the Hubble Legacy Archive, which is a collaboration between the Space Telescope Science Institute (STScI/NASA), the Space Telescope European Coordinating Facility (ST-ECF/ESA) and the Canadian Astronomy Data Centre (CADC/NRC/CSA).

 This research has made use of: 1) the VizieR catalogue access tool, CDS,
 Strasbourg, France. The original description of the VizieR service was
 published in Ochsenbein et al. 2000;   2)  the NASA/IPAC circumgalactic Database (NED) which is operated by the Jet Propulsion Laboratory, California Institute of Technology, under contract with the National Aeronautics and Space Administration; 

This research has made use of CosmoCalc \citealt{Wright2006} to obtain the scale kpc/" of each object.
This research (to correct Galactic extinction) has made use of the NASA/IPAC Extragalactic Database (NED), which is operated by the Jet Propulsion Laboratory, California Institute of Technology, under contract with the National Aeronautics and Space Administration.

We have extensively used the following software packages: TOPCAT (Taylor \citeyear{Taylor2005}), IDL Astronomy Library \citep{Landsman1993}, the Python routines of PyRAF and STSDAS ( STSDAS and PyRAF are products of the Space Telescope Science Institute, which is operated by AURA for NASA) and Matplotlib \citep{Hunter2007}.
This research made use of Astropy, a community-developed core Python package for Astronomy (Astropy Collaboration, \citeyear{Astropy2013}).

\newpage

\appendix
\label{appendix} 

\section{The point spread function}
\label{PSF}

\begin{table*}
\centering
\begin{tabular}{llllllllllllllll}
 \hline
Image         & Total & \multicolumn{2}{c}{Point source} &\vline &  &  Sersic 1 & & &\vline  & & Sersic 2 & &  &   \vline  &  Class  \\ 
	&   Mag & Mag	&  L.F. (\%)     & \vline  & Mag  & L.F. (\%)  & $r_e$ &  $n$   & \vline  & Mag  & L.F. (\%)  & $r_e$ &  $n$  &  \vline & \\   
 &  (1)  &  (2) & (3)  &   &  (4) &   (5)  &  (6) & (7)  &  & (8) &   (9) & (10) & (11) &  &  (12)      \\ \hline 
 Galaxy 1 &    &   & & &  &  &  \\ \hline
a) 	& 21.9	& 25.1 &	4.99	 & \vline  &  21.9 & 	95.01  &	4.8	& 1.0   &  \vline  & & & &  & \vline &  Disk-dominated \\
b) 	& 21.9	& 25.1  &	5.64	& \vline  &  22.0	& 94.36&	4.6	& 0.8 & \vline  &  & & & &  \vline &  Disk-dominated \\
c)  &	22.0	 & 25.5	& 3.74	& \vline  &  22.0& 	96.26	& 4.4	& 0.7  & \vline  &  & & &  & \vline &   Disk-dominated \\ \hline
 Galaxy 2 &    &   & & &  &  &  \\ \hline
a)   &	20.9	 	& 	 & &  \vline  & 20.9	& 100.00	&    2.7  &	0.9  & \vline  & & & &  & \vline &  Disk-dominated \\
b) 	&20.9	 & 	 & &   \vline  &	20.9	 & 100.00& 	2.9 &	1.0 & \vline   & & & &  & \vline &  Disk-dominated \\
c)  	& 21.0	 &	  & &  \vline  &	21.0	& 100.00	&  2.8	 & 0.9 &  \vline  & & & &  & \vline &  Disk-dominated \\ \hline
 Galaxy 3 &    &   & & &  &  &  \\ \hline
a) 	& 19.6	& 21.7 &	14.01 & \vline  & 	19.8	& 85.99	& 2.8	& 2.9 &  \vline  & & & &  & \vline &  Bulge-dominated  \\
b) &	19.5	 & 21.2	& 20.26  & \vline  &  	19.8	& 79.74 &	2.7	& 2.3  &  \vline  & & & &  & \vline &  Bulge-dominated  \\
c)   & 	19.6	 & 21.5	& 16.92	& \vline  &  19.8	& 83.08	& 2.4 &	2.0  &  \vline  & & & &  & \vline &  Bulge-dominated \\  \hline
 Galaxy 4 &    &   & & &  &  &  \\ \hline
a) &	20.7 &		& 	 &\vline  &    22.2	  & 25.33	& 1.7	& 1.4	&   \vline  & 21.0	& 74.67	& 6.2	& 0.3	& \vline &  Disk-dominated	 \\
b) &	20.4	   &    	 &      & \vline  &22.2	&  20.38	& 1.4	& 1.0 &   \vline  & 20.7	& 79.62  &	6.3 &	0.3 & \vline &  Disk-dominated	\\	
c)  &	20.4  & 		 &     & \vline  & 21.9  	  &  24.93	& 1.5	& 1.7&   \vline  	& 20.7  & 	75.07 &	6.4 &	0.2	& \vline &  Disk-dominated \\	\hline
 Galaxy 5 &    &   & & &  &  &  \\ \hline
a)   & 	20.9	 &    22.5	& 23.8	&   \vline  & 21.2	& 76.16  &	2.1   & 1.9	&  \vline  & & & &  & \vline &  Disk-dominated 	 \\			 	
b)   &	20.7	&    21.9  &	32.4  &	\vline  & 21.1	& 67.61	& 2.4	& 1.4	&  \vline  & & & &  & \vline &  Disk-dominated	\\
c)    & 	20.7	&  21.9 & 33.3	  & \vline  & 21.2 &	 66.67 &	2.6	& 1.0	&  \vline  & & & &  & \vline &  Disk-dominated \\		 \hline
\end{tabular}
 \caption{Test performed for several  normal galaxies  in the fields of three AGN of our WFPC2 sub-sample to investigate the impact of the PSF undersampling. The table shows the results of fitting with {\sc GALFIT} the host galaxies using a) the HLA image  b) the original raw unrotated image and c) this image  smoothed to fulfill Niquist sampling.  Col(1): Total magnitude Mag of the  {\sc GALFIT} model. Cols (2) and (3) Mag and light fraction  (L.F.) of the point source. Col(4) to (7) : Mag, L.F., effective radius $r_e$ in kpc  and index $n$ of first S\'ersic component . 
Col(8) to (11): Same for second S\'ersic component. Col(12) Galaxy classification following the criteria  in Sect. 3.2.4.  The classification  is consistent using a), b) and c) for all objects.}
 \label{tab:unders}
\end{table*}

\begin{table*}
\centering
\begin{tabular}{llllllllllllllll}
 \hline
Image         & Total & \multicolumn{2}{c}{Point source} &\vline &  &  Sersic 1 & & &\vline  & & Sersic 2 &   &  &  \vline  & Class \\ 
	& Mag	& Mag	& L.F. (\%)  &\vline & Mag & L.F. (\%)  & $r_e$ &  $n$  &\vline &  Mag & L.F. (\%) & $r_e$ &  $n$  & \vline  &  \\  
 &  (1)  &  (2) & (3)  &   &  (4) &   (5)  &  (6) & (7)  &  & (8) &   (9) & (10) & (11) &  &  (12)      \\ \hline 
& & & \multicolumn{3}{c}{SDSS J0748+39} & & & \\ \hline
a)   	& 18.7 &  21.3  &   9.45  &\vline & 18.6 & 90.55 & 9.4 & 4.3 &\vline &   &  && & \vline & Bulge-dominated   \\ 
b)   	& 18.6 &  21.3  &   8.63  &\vline& 18.5 & 91.37 &10.1& 4.5 &\vline &  & & & & \vline & Bulge-dominated      \\
c)  	& 18.7 &  22.2  &   4.14  &\vline& 18.5 & 95.86 &10.9& 5.3  &\vline &   & &  & & \vline & Bulge-dominated   \\ \hline
& & & \multicolumn{3}{c}{SDSS J0811+07} & & & \\ \hline
a)   	& 19.2 &     &   	    &\vline& 19.2 & 100.00     & 5.9 & 4.2  &\vline &   &  && & \vline & Bulge-dominated  \\
b)  	& 19.1 &     &        &\vline& 19.1 & 100.00     & 7.1 & 4.0  &\vline &  &  & & &    \vline & Bulge-dominated    \\
c)   	& 19.3 &     &        &\vline& 19.3 & 100.00     & 7.3 & 3.7 &    \vline  & &   &  & & \vline & Bulge-dominated    \\  \hline
& & & \multicolumn{3}{c}{SDSS J1726+60} & & & \\ \hline
a)    	& 19.8 &  20.6  & 47.06  &\vline& 20.3 & 52.94 & 2.6 & 1.4  & \vline  &  && & & \vline & Disk-dominated  \\ 
b)    	& 19.8 &  20.9  & 35.23  &\vline& 20.1 & 64.77 & 2.1 &  2.1 & \vline  &  && & & \vline & Bulge-dominated  \\
c) 	& 20.0 &  21.3  & 29.31  &\vline& 20.2 & 70.69 & 2.1 & 2.0  & \vline  &  && & & \vline & Bulge-dominated  \\ \hline
& & & \multicolumn{3}{c}{SDSS J1739+54} & & & \\ \hline
a)   	& 19.2 &  20.9  & 20.51  &\vline& 20.3 & 35.8 & 1.9 & 0.8  &\vline& 20.1 & 43.72 & 8.5 & 1.0 & \vline  & Disk-dominated \\ 
b)     	& 19.3 &  21.4  & 14.00  &\vline& 20.1 & 47.17 & 1.6 & 1.3  &\vline& 20.2 & 38.83 & 9.6 & 1.0 & \vline  & Disk-dominated  \\ 
c) 	& 19.5 &  23.0  &  3.38   &\vline& 20.1 & 55.47 & 1.3 & 1.0  &\vline& 20.3 & 41.15 & 8.7 & 0.8 & \vline  & Disk-dominated \\ \hline   
 \end{tabular}
 \caption{Same as Table A2 with tests performed to four AGN in the WFPC2 sample. The classification of the objects is in general consistent using a), b) and c).}
 \label{tab:unders2}   
\end{table*}

The final model image produced by GALFIT for a given quasar  is constructed by convolving a  model image of the host galaxy with a PSF. Thus, the success of the method relies on knowing an accurate PSF. This is particularly important in studies of type 1 (unobscured)  AGN, where the central unresolved source can dominate entirely the flux in the central regions and the PSF wings can contaminate severely at large radii.  Small variations of the PSF can lead to large variations on the inferred host galaxy properties (Kim et al. 2008, \citealt{Kim2008} hereafter). 

  Although the central source in type 2 AGN is obscured and the PSF effects are less severe than in the  type 1 counterparts,  a good quality PSF is required. The presence of an unresolved nuclear source cannot be discarded.
   Scattered  AGN light,  nebular continuum, emission lines and nuclear star clusters may all contribute to the nuclear emission (e.g. \citealt{Ballcells2007}, \citealt{Bessiere2017}). The PSF profile is so steep that it can have a significant impact on the central shape and the inferred parametric properties of a galaxy even for  low contributions   to the total flux  ($\la$10\%).
 
 \vspace{0.2cm}
 \centerline{\it Colour, temporal and spatial variations of the PSF}
 \vspace{0.2cm}

The HST PSF are known to vary  depending on  position on the detector, time and  spectral colour. 
Other authors (see in particular \citealt{Kim2008}) have performed careful investigations of such dependence.  Based on these works, we have assumed that the influence of the PSF colour is comparatively negligible. The ideal procedure to obtain a good quality PSF would then be to select one  star or a combination of several in each AGN field close to the target ($\la$100 pixels for the WFPC2, \citealt{Kim2008}). However,  appropriate stars (i.e. not saturated and with adequate S/N to trace  the core and the wings) at $\la$100 pixels are unavailable for most of the objects. 

Since \citet{Kim2008} found that spatial variations of the PSF are significantly smaller than   temporal variations, our aim has been  to minimize temporal variations. The HST WFPC2 observations of our sample were concentrated around two different epochs: April to June  2007  and November 2008. The ACS observations were performed between August and December 2006.  

We distributed the WFPC2 subsample in two groups (2007 and 2008) according to the two  observing epochs (see Table \ref{Table:0}). We then  generated two libraries of PSF, one for each epoch. Each library contains   2D  images of non-saturated stars with well detected wings selected in the images of different AGN targets. A third PSF library was built for the ACS sample. 

 For each epoch/instrument, we combined several stars of the field in the corresponding library normalizing the flux to unity and weighting by the S/N ratio \citep{Inskip2010}.

\vspace{0.2cm}
 \centerline{\it Undersampling of the PSF}
\vspace{0.2cm}

  Another aspect to take into account is that the HST PSF is undersampled in the WFPC2 images, for which the nominal FWHM is $\sim$1.5 pixels.  The main impact  is that it is not possible to preserve the original shape of the PSF when shifting by a fraction of a pixel \citep{Kim2008, Peng2002}. As explained by \cite{Kim2008}, the subpixel interpolation can change significantly both the width and the amplitude of the unresolved flux, while the wings of the PSF, which are much better sampled, do not change much. 

 The HLA images of our sample have a  FWHM$\sim$2.1-2.5 pixels  depending on the object, which is consistent with Nyquist-sampling. The  images in the HLA are drizzled, i.e. the original pixels were mapped onto an output rotated frame, where a single pixel from the detector might be spread over more than one output pixel. 
 This process can spread the PSF out to a larger value than the original PSF in the unrotated images. 
  
In order to investigate the impact of an undersampled PSF, we applied a diversity of tests to  5 random normal galaxies and 4 random  AGN hosts of our sample,   observed in the two WFPC2 epochs and spanning morphological diversity. For these purposes, we consider ``normal''  galaxies, those  different to the AGN targets with no obvious evidence for  nuclear activity (i.e. a prominent central point source) and  morphological distortions.  The AGN and normal galaxies  where fitted with {\sc GALFIT} in 3 different images corresponding to the same field of view: a)  the HLA images  using combined field stars to produce the PSF (see above),  b)  the original, unrotated calibrated images using also combined field stars for the PSF,  c)  the original, unrotated calibrated images convolved with a Gaussian whose FWHM ensures achieving Nyquist-sampling  \citep{Kim2008}. The PSF was convolved in the exact same way.\footnote{This method was also attempted using  PSF created with TinyTim \citep{Krist1995}, but the PSF we obtained from the data yielded better results}

We find that a), b) and c) produce consistent results for normal  (Table \ref{tab:unders}) and AGN host galaxies (Table \ref{tab:unders2})   in the sense that the same number of structural components  are required for a given object. The best fits are also in general consistent in terms of the $r_e$ and $n$  of  each  component.  The relative contribution to the total light of a given structural component can vary up to $\sim$10\%   at most for a given structural component for normal galaxies and up to $\sim$20\% for AGN hosts. The final classification of the galaxy is always consistent in a), b) and c) for normal galaxies and, in general, for AGN hosts. SDSS J1726+60 is an exception in the AGN group, although this is not surprising, since it has an intermediate $n$ value between disk-dominated and bulge-dominated systems, so that the final classification is strongly sensitive to small $n$ uncertainties.

Our conclusion is that the effects of undersampling for the WFPC2 do not have a significant impact on the structural parametrization of the galaxies.

\section{Brief description of related studies}
\label{otherworks}

We presented in Sect. \ref{Analogous:Studies} a  comparison of the results of our parametric classification with related studies.
We describe  here very briefly the main properties of the samples, data,  methodology and classficiation criteria presented in these works.

$\bullet$ \citet{Dunlop2003}  carried out a seminal work based on R band HST/WFPC2  images  of the host galaxies of 13 type 1 radio-quiet  (RQQ) and 10 type 1  radio-loud quasars  (RLQ) at 0.11$< z <$0.26 with nuclear absolute magnitudes in the range  -19.7$\le M_ {\rm R} \le $-25.7. They fitted the host galaxies  with one or two S\'ersic components. All  RLQ and 9 RQQ showed no evidence for any disk component ($n\sim$1) and were classified as ellipticals as a result, based on the galaxy $n\ga$4 profile. Only 4 RQQ were best fitted with a combination of a disk and a bulge, two of which are dominated by the spheroid. Therefore, the analysis by \citet{Dunlop2003} results on 10/10 or $100\%_{-16}$ RLQ and 11/13 or $85\%^{+6}_{-15}$ RQQ are spheroidal or bulge-dominated. While only $15\%^{+15}_{-6}$ of RQQ are disk-dominated.

$\bullet$ \citet{Floyd2004}  studied a sample of 17 QSO1 (10 radio-quiet and 7 radio-loud) at  0.29$< z<$0.43, spanning a range of  absolute magnitudes -27.7$\le M_ {\rm V} \le $-24.4, using HST/WFPC2 images and the F814W or F791W filters. 
 Accounting also for the point source central function, the authors fitted the host galaxy surface brightness profiles using a single S\'ersic component. When the index is left free, they find 13/17 ($76\%^{+8}_{-13}$) with $n\ga$2.5 that they classify as ellipticals (see their Table 4), 3/17 ($18\%^{+11}_{-7}$)  with $n=$0.75-1.04 that they classify as disks galaxies and 1/17 ($6\%^{+11}_{-3}$) is an intermediate case with $n=$1.8. 

$\bullet$  \citet{Falomo2014}  studied the galaxy types and morphologies of 416 QSO1 at z$<$0.5 with $M_{\rm i}<$-22 using i-band SDSS images in the Stripe82 region that are significantly deeper than standard SDSS data. Most are radio-quiet. Galaxies were well resolved in 316 objects.  For the classification of the host morphologies they combine the visual inspection of the images and the fits of the light profiles  with a PSF and a single S\'ersic component  using the Astronomical Image Decomposition Analysis (AIDA; \citealt{Uslenghi2008}). They consider two types of morphology: exponential disk and de Vaucouleurs profile. They find that about 113  objects ($37\%^{+3}_{-3}$) are dominated by the bulge, 129 ($42\%^{+3}_{-3}$) have a conspicuous disk structure and 64  ($21\%^{+3}_{-3}$) exhibit a mixed bulge plus disk structure.

$\bullet$ \citet{Cales2011} studied a sample of 29 post starburst QSO1 at 0.25$< z <$0.45 with  -24.0$\le M_r\le$-22.1 using HST/ACS-F606W  images.  No information is quoted on radio-loudness. The authors classify the galaxies visually. In this way they identify  an equal number of spiral (13/29, $45\%^{+8}_{-10}$) and early-type (13/29, $45\%^{+8}_{-10}$) hosts, with the remaining three hosts having indeterminate classifications (3/29, $10\%^{+8}_{-4}$). They also  parametrized the galaxies with {\sc GALFIT}, selecting the best fit for each object aided by the prior visual classification. They found that galaxies visually classified as early types are fitted with a single S\'ersic component ($n\ga$2 in most cases) and that most  galaxies visually classified as spirals or probable spirals are fitted with two S\'ersic components with fixed index: $n$=4 for the bulge and $n$=1 for the disk.

Because the information in  \citet{Cales2011} is not enough to identify all bulge-dominated systems, we have  applied  our parametric method to classify their sample (Table \ref{compa-merger}). For this, we have used the parameter values for each structural component the authors isolate  in their fits. Based on their highly-disturbed morphology,  we classify 4  objects  as disturbed (or ``Multiple Systems").   We find  13 spheroidal galaxies, 5 bulge-dominated galaxies, 3  disk-like galaxies, 3 disk-dominated galaxies and 1 object for which the bulge and the disk have similar contribution. The final classification is:   $21\%^{+8}_{-7}$ disk-dominated, $62\%^{+4}_{-14}$ bulge- dominated, $3\%^{+7}_{-1.2}$ B+D galaxies and $14\%^{+8}_{-5}$ disturbed systems.

$\bullet$ \citet{Greene2009} studied a sample of 15 SDSS QSO2 at 0.1$< z <$0.45 with 8.7$\la lO3 \la$9.3. No information is provided on the radio-loudness, but a fraction of $\sim$15$\pm$5\% can be expected to be radio-loud \citep{Lal2010}. They parametrized the host galaxies using {\sc GALFIT} based on ground based $r$  band images (except for 3 objects observed with the $g$ and $i$ filters).  The images were obtained with the  Low Dispersion Survey Spectrograph (LDSS3) at the 6.5m Clay-Magellan telescope. The authors were mostly interested on the spheroidal components. For this reason, they only introduce a disk component ($n$=1)  in the galaxy model when visible in the images. In this way, they identify 4 disk galaxies (with the bulge dominating in one of them) while seven  consist of a single spheroidal component. 4 highly-disturbed objects could not be fitted successfully.
Therefore, the final classification is  8/15 ($53\%^{+9}_{-16}$) are bulge-dominated, 3/15 ($ 20\%^{+12}_{-8}$) are disk-dominated and 4/15 ($ 27\%^{+12}_{-10} $ ) are highly-disturbed systems.

$\bullet$ \citet{Wylezalek2016} analyzed HST/ACS FR914M optical images of 20 luminous  ($lO3\ge$9.0) radio-quiet SDSS QSO2 at 0.2$< z <$0.6. They fitted the host galaxies with {\sc GALFIT} using one or two S\'ersic components. For single components, they consider that a galaxy is disk-dominated when $n\le$1 or bulge-dominated for $n\ge$1. When two components are isolated in the fits, they classify a galaxy as disk-dominated when the $n$ of the brighter (or primary) component is $n_ {\rm pri}\le$ 1 and bulge-dominated when  $n_ {\rm pri}\ge$1. According to this method, they find that all but 2 QSO2 are bulge-dominated  ($90\%^{+4}_{-11}$), while the remaining 2 are disk-dominated ($10\%^{+11}_{-4}$).

$\bullet$ \citet{Inskip2010} studied the parametric properties of a sample of 41 radio galaxies at 0.03$\leq z \leq$ 0.5: 17 narrow line radio galaxies (NLRG), 12 broad line radio galaxies (BLRG) and 13 weak line radio galaxies (WLRG). They used ground based K-band images obtained with the instrument/telescope combinations UFTI/UKIRT, ISAAC/VLT and, most them, with SOFI/NTT.  Here we compare  with the NLRG sample: although powerful radio sources, if they were classified based only  on their optical emission line spectroscopic properties \citep{Zakamska2003} they would be classified as   QSO2.

The authors used {\sc GALFIT} for their analysis in 16 NLRG (one of them was too faint to model). They fit most  host galaxies with a single S\'ersic component, including also a  point source in a significant  fraction of objects (see Sect. \ref{psf}).   In general, most galaxies (14/16  or $88\%^{+1}_{-17}$) are fitted with a $n$=4 or $n$=6 S\'ersic, which the classify as ``bulges".  1/16 or $6\%^{+11}_{-3}$ is fitted with a $n$=2 S\'ersic, which they classify as ``disk". Finally, 1/16 or $6\%^{+11}_{-3}$  consists of a disk and bulge, which they classify as ``mixed".

\end{document}